\documentclass{article}
\usepackage[dvipdfmx]{graphicx}
\usepackage{amsmath}
\usepackage{amssymb}
\usepackage{authblk}
\usepackage{mathrsfs}
\usepackage{slashed}
\usepackage{color}
\usepackage{cite}
\usepackage{here}
\usepackage{subcaption}

\makeatletter
 
 \@addtoreset{equation}{section}
 \makeatother

\newcommand{\be}{\begin{equation}}
\newcommand{\ee}{\end{equation}}

\newcommand{\bea}{\begin{equnarray}}
\newcommand{\eea}{\end{eqnarray}}
\newcommand{\beann}{\begin{equnarray*}}
\newcommand{\eeann}{\end{eqnarray*}}
\newcommand{\nn}{\nonumber}
\newcommand{\ba}{\begin{array}}
\newcommand{\ea}{\end{array}}

\newcommand{\bs}{\boldsymbol}

\newcommand{\bse}{\bs{e}}

\newcommand{\lrangle}[1]{\left\langle{#1}\right\rangle}
\newcommand{\lrpar}[1]{\left({#1}\right)}
\newcommand{\sgn}{{\rm sgn}\,}
\newcommand{\N}{\mathbb N}
\newcommand{\bfx}{\mathbf{x}}

\DeclareMathOperator{\diag}{diag}

\DeclareMathOperator{\Tr}{Tr}

\title{Kazakov-Migdal model on the Graph and Ihara Zeta Function}
\author[1]{So Matsuura\thanks{s.matsu@phys-h.keio.ac.jp}}
\author[2]{Kazutoshi Ohta\thanks{kohta@law.meijigakuin.ac.jp}}
\affil[1]{\it Hiyoshi Departments of Physics,
and Research and Education Center for Natural Sciences,
Keio University, 4-1-1 Hiyoshi, Yokohama, Kanagawa 223-8521, Japan}
\affil[2]{\it Institute of Physics, Meiji Gakuin University, Yokohama, Kanagawa 244-8539, Japan}

\date{}

\begin{document}
\maketitle

\vspace*{2cm}

\begin{center}
{\bf Abstract}
\end{center}

We propose the Kazakov-Migdal model on graphs and show that, when the parameters of this model are
appropriately tuned, the partition function is represented by the unitary matrix integral of an extended 
Ihara zeta function, which has a series expansion by all non-collapsing Wilson loops with their lengths as 
weights. The partition function of the model is expressed in two different ways according to the order of 
integration. A specific unitary matrix integral can be performed at any finite $N$ thanks to this duality. 
We exactly evaluate the partition function of the parameter-tuned Kazakov-Migdal model on an arbitrary 
graph in the large $N$ limit and show that it is expressed by the infinite product of the Ihara zeta 
functions of the graph.

\newpage

\section{Introduction}
\label{sec:Introduction}

The large $N$ limit in gauge theory has so far led to many insights not only
in gauge theory but also in string theory and has resulted in AdS/CFT correspondence.
In this connection, for various types of matrix models, analysis in the large $N$ approximation
has yielded important results in a wide range of physics and mathematics.
The matrix models have appeared from various points of view
such as toy models of QCD, large $N$ reduction of  gauge theories, 
the exact partition functions of supersymmetric gauge theories, 
and more.

Kazakov and Migdal proposed a lattice gauge model, which induces the
effective gauge field action including multiple Wilson loops \cite{kazakov1993induced}.
The Kazakov-Migdal (KM) model consists of a scalar field in the adjoint representation
on sites and gauge field (unitary matrix) on links of the lattice.
The action is inspired by the well known Harish-Chandra–Itzykson–Zuber (HCIZ) integral
over a unitary group with two Hermite matrices
\cite{10.2307/2372387,Itzykson:1979fi}.
In the large $N$ limit, the model is studied by using contemporary matrix model techniques.

Although the original KM model is defined on a square lattice in various dimensions, 
we can extend it to a model on a generic graph, which describes a kind of discretization of space-time 
as we show in this paper. 
By adjusting the coupling constants in the extended KM model on the graph,
the partition function of the model reduces to
a generating function of the ``non-collapsing'' Wilson loops corresponding to reduced cycles in graph theory, 
thanks to the important results of the {\it Ihara zeta function}
\cite{Ihara:original,MR607504,sunada1986functions}
(see also \cite{terras_2010} for a recent review). 

The Ihara zeta function 
associated with a finite graph 
is an analog of the Riemann zeta function, 
which plays very important role in the number theory.
The Ihara zeta function and its related theorems are very beautiful and meaningful and have been applied mainly to mathematics, 
but have not been applied much in high energy physics, 
except for applications to quiver gauge theory \cite{He:2011ge,Zhou:2015tia,Ohta:2020ygi} 
and discretized (supersymmetric) gauge theory
\cite{Kan:2004kz,Kan:2009tu,Kan:2013mra,Matsuura:2014kha,Matsuura:2014nga,Kamata:2016xmu,Ohta:2021tmk}.

In this paper, we give an essential relationship between the Kazakov-Migdal model on the graph and the Ihara zeta function. 
We show that, by appropriately tuning the parameters of the model, the partition function is expressed as the unitary matrix integral of an extended Ihara zeta function, 
which has a series expansion by all non-collapsing Wilson loops with their lengths as weights. 
We will show that, in the large $N$ limit,
the unitary matrices (and also the scalar fields) can be integrated out
completely and the partition function is given by an infinite product of the Ihara zeta function.
This means that we can exactly solve the KM model on the graph in the large $N$ limit.

The organization of this paper is as follows:
In the next section, we give a brief review on the Ihara zeta function
for unfamiliar readers. It contains definitions of various terminologies of graph theory.
For later discussions, we also extend the Ihara zeta function to include matrices on the edges.
In Sect.~3, we define the KM model on a generic graph.
We show that the partition function of the graph KM model is expressed by
the integral of the extended Ihara zeta function over the unitary matrices, after integrating the scalar
field first.
This model has a different picture depending on the order of integration.
If we integrate the unitary matrices first by using the HCIZ integral formula,
we obtain a dual integral expression over the eigenvalues of the Hermitian scalar fields.
As a byproduct, we can obtain the exact value of the integral
$\int dU \exp\lrpar{\sum_{n=1}^\infty \frac{q^n}{n} |\Tr U^n|^2}$
at finite $N$ by using the dual description in cycle graphs. 
In Sect.~4, we perform the integral of the extended Ihara zeta function over the unitary matrices.
For finite $N$, this integral is generally difficult except for the cycle graphs.
However, in the large $N$ limit, due to the cluster decomposing nature of the expectation value
of the Wilson loops, the integrals can be performed exactly. Our main result is that
the partition function of our model is given by an infinite product of the Ihara zeta function.
The last section is devoted to the conclusion and discussion.
In Appendix \ref{Ihara examples}, we show some examples of the Ihara zeta function. 
In Appendix \ref{app:proof}, we give a proof of identities associated with symmetric polynomials used in this paper. 
In Appendix \ref{app:q1}, we directly carry out the unitary matrix integral appearing in 
Sect.~3 for small $N$ using Cauchy's integral formula.

\section{Graph Theory and the Ihara Zeta Function}
\label{sec:Ihara zeta}
In this section,
we give definitions of some objects appeared in graph theory
and a brief review of the Ihara zeta function.
We then propose an extension of the Ihara zeta function using matrices.

\subsection{Ihara zeta function}

A graph $G$ consists of vertices connecting with edges.
We denote the set of vertices and edges as
$V$ and $E$, respectively. The numbers of the vertices and edges
(the numbers of the elements in $V$ and $E$) are also denoted by
$n_V=|V|$ and $n_E=|E|$.
We here assume the graph is connected and directed. So each edge has
the direction.
A directed edge $e=\langle u,v\rangle$ connecting the vertices $u,v\in V$ has
the ``source'' (starting point) $u=s(e)$ and ``target'' (end point) $v=t(e)$. 
$e^{-1}=\langle v,u\rangle$ is called the inverse of $e$.
We denote the set of the edges with their inverses as
\begin{equation}
E_D = \{\bse_a|a=1,\cdots,2n_E\} \equiv \{e_1,\cdots,e_{n_E},e^{-1}_1,\cdots,e^{-1}_{n_E}\}\,.
\nn
\end{equation}
In other words, $E_D$ is the direct sum of the set of edges $E$ and the set of the inverse of edges $E^{-1}$; $E_D=E\oplus E^{-1}$.

A path $P=\bse_1 \bse_2 \cdots \bse_\ell$ $(\bse_a \in E_D)$ is a sequence of
the edges which satisfies $t(\bse_a)=s(\bse_{a+1})$ $(a=1,\cdots,\ell-1)$,
where $\ell$ is called the length of the path $P$,
which is denoted as $\ell(P)$.
If two paths $P=\bse_1\cdots\bse_\ell$ and $P'=\bse'_1\cdots\bse'_{\ell'}$ satisfy
$t(\bse_k)=s(\bse'_1)$, we can construct a new path of length $\ell+\ell'$
by connecting them as
$PP'\equiv \bse_1\cdots \bse_\ell \bse'_1,\cdots,\bse'_{\ell'}$.
A backtracking of $P$ is a part of $P$ which satisfies $\bse_{a+1}^{-1} = \bse_a$.

When the path $P$ satisfies $s(\bse_1)=t(\bse_\ell)$,
$P$ is called a cycle of length $\ell$.
A cycle $C= \bse_1\bse_2\cdots\bse_\ell$ is called tailless when $\bse_\ell^{-1}\ne \bse_1$,
which is equivalent to that $C^2$ has no backtracking.
A cycle $C$ is called reduced when $C$ has no backtracking nor tail.
A reduced cycle $C$ is called primitive when $C$ satisfies $C \ne B^r$ $(r\ge 2)$
for any reduced cycle $B$.

We call two cycles $C=\bse_1\cdots\bse_\ell$ and $C'=\bse'_1\cdots\bse'_{\ell'}$ equivalent when $\bse_a = \bse'_{a+r}$ for an integer $r$,
namely the difference between two cycles is only the starting point.
Using the equivalence of the cycles $C\sim C'$, we can define the equivalence class
of the cycles, which is denoted by $[C]$;
\be
[C] = \{
\bse_1\bse_2\cdots\bse_\ell,\ 
\bse_2\bse_3\cdots\bse_1,\ \ldots\ ,\ 
\bse_\ell\bse_1\cdots\bse_{\ell-1}
\}.
\nn
\ee

The Ihara zeta function
corresponding to a graph $G$ is defined as \cite{Ihara:original,sunada1986functions}
\begin{align}
    \zeta_G(q) \equiv \prod_{[C]:\text{primitive}} \frac{1}{1-q^{\ell(C)}}\,,
    \label{eq:Ihara zeta}
\end{align}
where $q\in \mathbb{C}$ is assumed to have sufficiently small $|q|$,
and $[C]$ runs the equivalence classes of the primitive cycles on the graph $G$.

Noting that, if $C$ is a primitive cycle, $C^{-1}$ is also a primitive cycle with the same length as $C$,
a set of the equivalence classes of the primitive cycles 
of length $\ell$ can be decomposed%
\footnote{
This decomposition is not unique but the number of the elements
of $\Pi_\ell^+$ is unique.
It is enough for the following discussion.
}
into 
$\Pi_\ell^+ \sqcup \Pi_\ell^-$, where 
$\Pi_\ell^-$ is the set of the inverse of the elements in $\Pi_\ell^+$ such that 
$\Pi_\ell^-\equiv \left\{C^{-1}|C\in \Pi_\ell^+\right\}$.
Since these sets
$\Pi_\ell^+$ and $\Pi_\ell^-$ have the same number of the elements by definition; $|\Pi_\ell^+|=|\Pi_\ell^-|$,
the Ihara zeta function \eqref{eq:Ihara zeta} can also be rewritten as
\begin{align}
  \zeta_G(q) = \prod_{\ell=1}^\infty \frac{1}{(1-q^\ell)^{2|\Pi^+_\ell|}}\,.
  \label{eq:Ihara zeta with Nk}
\end{align}
In the following, we call the element of $\Pi_\ell^+$ as
the equivalence class of {\em chiral primitive cycles} of length $\ell$.

For later use, we give another expression of the Ihara zeta function.
By using $(1-x)^{-1}=\exp(\sum_{m=1}^\infty \frac{x^m}{m})$,
we can rewrite \eqref{eq:Ihara zeta with Nk} as
\begin{align}
  \zeta_G(q) = \exp\lrpar{\sum_{\ell=1}^\infty \sum_{m=1}^\infty
  \frac{2|\Pi^+_\ell|}{m} q^{\ell m}}\,.
  \nn
\end{align}
Since any reduced cycle is expressed as a positive power of a primitive cycle $C$ and 
$[C^n]$ $(n\in{\mathbb N})$ is consist of $\ell(C)$ different reduced cycles,
$2\ell |\Pi^+_\ell|$ is the number of the reduced cycles of length $\ell$.
Therefore we can regard the Ihara zeta function \eqref{eq:Ihara zeta}
as the generating function of the number of the reduced cycles:
\begin{align}
\zeta_G(q) = \exp\left( \sum_{k=1}^\infty \frac{N_k}{k} q^k \right)\,,
    \label{eq:Ihara zeta as generating function}
\end{align}
where $N_k$ is the number of the reduced cycles of length $k$.

\subsection{Ihara zeta function via the edge adjacency matrix}

It is remarkable that the Ihara zeta function is expressed as an inverse of a polynomial of $q$.
To show it, we define a $2n_E\times 2n_E$ matrix $W$
called the edge adjacency matrix of $G$ as \cite{hashimoto1989zeta,hashimoto1990zeta}
\begin{align}
    W_{\bse\bse'} = \begin{cases}
      1 & {\rm if}\ t(\bse) = s(\bse')\ {\rm and}\ \bse'^{-1}\ne \bse \\
      0 & {\rm others}
    \end{cases}\,,
    \label{eq:edge matrix}
\end{align}
where $\bse, \bse' \in E_D$.

It is obvious from the construction
that the element $(W^k)_{\bse\bse'}$ is the number of
the paths of length $k$
whose source and target are $s(\bse)$ and $s(\bse')$, respectively,
but the last edge is {\em not} $\bse'^{-1}$.
In particular, the diagonal element of a power of $W$, $(W^k)_{\bse\bse}$,
is the number of the reduced cycles of length $n$ which start from $\bse$.
Therefore, we see
\begin{align}
    \sum_{k=1}^\infty \frac{N_k}{k} q^k
    = \sum_{k=1}^\infty \frac{q^k}{k}\Tr W^k
    = -\Tr \log\left( \bs{1}_{2n_E} - q W\right) \,.
\nn
\end{align}
Comparing this expression to \eqref{eq:Ihara zeta as generating function}
we obtain \cite{hashimoto1989zeta,hashimoto1990zeta,bass1992ihara}
\begin{align}
    \zeta_G(q) = 
    \det\Bigl(
      \bs{1}_{2n_E} - q W
    \Bigr)^{-1}\,,
    \label{eq:determinant expression2}
\end{align}
which is the inverse of a polynomial of $q$ as announced.

\subsection{Ihara's theorem}

We define an $n_V\times n_V$ matrix $A$ called the adjacency matrix of $G$ as
\begin{align}
    A_{vv'} = \sum_{e\in E}\lrpar{
    \delta_{v,s(e)}\delta_{v',t(e)} + \delta_{v,t(e)}\delta_{v',s(e)}}\,,
    \label{eq:adjacency matrix}
\end{align}
where $v,v'\in V$,
which is the number of $\lrangle{v,v'}$ in $E_D$.
We define the degree of the vertex $v$ as
\[
  \deg v \equiv \sum_{v'\in V} A_{vv'},
\]
which is the summation of the number of edges which start from or end on the vertex $v$.
We then define the degree matrix $D$ of the graph $G$ as
\begin{align}
    D_{vv'} = \deg v \, \delta_{vv'}\,.
    \nn
\end{align}
In \cite{Ihara:original}, it is shown that the Ihara zeta function is expressed by
a rational function,
\begin{align}
    \zeta_G(q) = (1-q^2)^{n_V-n_E}\det\left(
      \bs{1}_{n_V} - qA + q^2 (D-\bs{1}_{n_V}) \right)^{-1}\,.
    \label{eq:determinant expression1}
\end{align}
This result is called Ihara's theorem.
Some examples of Ihara's theorem are given in Appendix~\ref{Ihara examples}.

Instead of proving this theorem,
we will extend the Ihara zeta function
and prove a similar theorem with respect to the extended function.

\subsection{Extended Ihara zeta function}

Suppose $X_{e}$ $(e\in E)$ are invertible matrices of size $K$ living on each edge.
Using these matrices, we extend the Ihara zeta function as
\begin{align}
    \zeta_G(q;X) \equiv
    \det\Bigl(
      \bs{1}_{2n_E}\otimes\bs{1}_{K} - q W_X
    \Bigr)^{-1}\,,
    \label{eq:extended Ihara1}
\end{align}
where $W_X$ is an extension of the edge adjacency matrix \eqref{eq:edge matrix}
whose elements are matrices of size $K$:
\begin{align}
    (W_X)_{\bse\bse'} = \begin{cases}
      X_{e} & {\rm if}\ \bse=e\in E, \  t(e) = s(\bse'),\ {\rm and}\ \bse'^{-1}\ne e \\
      X_{e}^{-1} & {\rm if}\ \bse=e^{-1}\in E^{-1},\ s(e) = s(\bse'),\ {\rm and}\ \bse'^{-1}\ne e^{-1} \\
      0 & {\rm others}
    \end{cases}\,.
    \label{eq:edge matrix X}
\end{align}
Note that (\ref{eq:extended Ihara1}) reduces to the weighted Ihara zeta function discussed in \cite{mizuno2004weighted} when $K=1$ (weighted by complex scalar
numbers).
From the construction, the diagonal element of $n$'s power of $W_U^n$, $(W_U^n)_{\bse\bse}$,
is a sum of the product of $X_e$ along a reduced cycle $C$ of length $n$
which is denoted by $P_{C}(X)$.
For example, if $C=e_1e_2^{-1}e_3$, $P_{C}(X)=X_{e_1}X_{e_2}^{-1}X_{e_3}$.
Therefore, we can rewrite the extended Ihara zeta function \eqref{eq:extended Ihara1} as
\begin{align}
  \zeta_G(q;X) = \exp\lrpar{
    \sum_{k=1}^\infty \frac{1}{k}\sum_{C\in R_k}\Tr P_{C}(X)
    }\,,
    \label{eq:extended Ihara P}
\end{align}
where $R_k$ is the set of reduced cycles of length $k$.

In the following, we show that the extended Ihara zeta function
\eqref{eq:extended Ihara1} can be expressed as
\begin{align}
    \zeta_G(q;X) = (1-q^2)^{(n_V-n_E)K}\det\Bigl(
      \bs{1}_{n_V}\otimes\bs{1}_{K} - qA_X
      + q^2(D-\bs{1}_{n_V})\otimes\bs{1}_K \Bigr)^{-1}\,,
    \label{eq:extended Ihara2}
\end{align}
where $A_X$ is an extension of the adjacency matrix \eqref{eq:adjacency matrix} defined by
\begin{align}
    (A_X)_{vv'} =
    \sum_{e\in E} \lrpar{
    X_e\, \delta_{\lrangle{v,v'},e}
    +X_e^{-1}\, \delta_{\lrangle{v',v}, e}
    }\,,
    \label{eq:adjacency matrix X}
\end{align}
whose elements are matrices of size $K$ as well as $W_X$.
Since this relation holds for $K=1$ and $X=1$,
the original Ihara's theorem is a corollary.
The following proof is parallel to the one given
in \cite{bass1992ihara} for the original Ihara zeta function:

Firstly, we define matrices $S_X$ and $T_X$ of size $n_E\times n_V$
whose elements are matrices of size $K$;
\begin{align}
    (S_X)_{ev} \equiv \begin{cases}
      X_e^{-1} & {\rm if}\ v=s(e) \\
      0 & {\rm others}
    \end{cases}\,,\quad
    (T_X)_{ev} \equiv \begin{cases}
      X_e & {\rm if}\ v=t(e) \\
      0 & {\rm others}
    \end{cases}\,.
    \label{S and T}
\end{align}
We also define $S\equiv (S_X)|_{X=1}$ and $T\equiv (T_X)|_{X=1}$.
Then we can easily show
\begin{align}
    S^T T_X  + T^T S_X = A_X, \quad
    S^T S + T^T T = D\otimes \bs{1}_K\,.
    \nn
\end{align}

Using these matrices, we define
\begin{align}
    L&\equiv \begin{pmatrix}
        \bs{1}_{n_V}\otimes \bs{1}_{K} & qS^T & q T^T \\
        -q S+T_X & (1-q^2) \bs{1}_{n_E}\otimes \bs{1}_{K}  & 0 \\
        -q T+S_X & 0 & (1-q^2) \bs{1}_{n_E}\otimes \bs{1}_{K}
    \end{pmatrix}\,, \nn \\
    M&\equiv \begin{pmatrix}
        (1-q^2) \bs{1}_{n_V}\otimes \bs{1}_{K} & 0 & 0 \\
        qS-T_X & \bs{1}_{n_E}\otimes \bs{1}_{K} & 0 \\
        qT-S_X & 0 & \bs{1}_{n_E}\otimes \bs{1}_{K}
    \end{pmatrix}\,. \nn
\end{align}
After a straightforward calculation, we obtain
\begin{align}
    LM &= \begin{pmatrix}
       \bs{1}_{n_V}\otimes\bs{1}_{K} - qA_X
      + q^2(D-\bs{1}_{n_V}) \otimes \bs{1}_{K} & qS^T & qT^T \\
        0 & (1-q^2) \bs{1}_{n_E} \otimes \bs{1}_{K} & 0 \\
        0 & 0 & (1-q^2) \bs{1}_{n_E} \otimes \bs{1}_{K}
    \end{pmatrix}\,,  \nn\\
    ML&= \begin{pmatrix}
        (1-q^2)\bs{1}_{n_V} \otimes \bs{1}_{K} &  q(1-q^2)S^T \hspace{1cm} q(1-q^2)T^T \\
        \begin{array}{c} 0 \\ 0 \end{array} &
        (\bs{1}_{2n_E} \otimes \bs{1}_{K}-q J_X)  (\bs{1}_{2n_E} \otimes \bs{1}_{K}-qW_X)
    \end{pmatrix}\,,
    \nn
\end{align}
where
\begin{align}
    J_X \equiv \begin{pmatrix}
      0 & \begin{matrix}X_{e_1} & \cdots & 0 \\ \vdots & \ddots & \vdots \\ 0 & \cdots & X_{e_{n_E}}\end{matrix} \\
      \begin{matrix}X_{e_1}^{-1} & \cdots & 0 \\ \vdots & \ddots & \vdots \\ 0 & \cdots & X_{e_{n_E}}^{-1}\end{matrix}
      & 0
    \end{pmatrix}\,.
    \nn
\end{align}
Using $\det(LM)=\det(ML)$ and
${\rm det}\left( \bs{1}_{2n_E} \otimes \bs{1}_{K}-q J_X \right) = (1-q^2)^{Kn_E}$,
we find
\begin{multline}
\qquad
(1-q^2)^{2 K n_E}\det\left(\bs{1}_{n_V}\otimes\bs{1}_{K} - qA_X
      + q^2(D-\bs{1}_{n_V}) \otimes \bs{1}_{K} \right)\\
=(1-q^2)^{K (n_V+n_E)}\det\left(\bs{1}_{2n_E} \otimes \bs{1}_{K}-qW_X\right)\, ,
\qquad
\nn
\end{multline}
and then conclude (\ref{eq:extended Ihara2}).

\section{Graph Kazakov-Migdal model}
\label{sec:model}

In this section, we propose a matrix integral that adds up all the Wilson loops on the graph with their lengths as weights.
This integral is obtained by appropriately adjusting the parameters of the KM model on the graph.
As an application, a certain integral over a unitary matrix of finite size is evaluated exactly through elementary linear algebra.

\subsection{Kazakov-Migdal model on the graph}

The KM model \cite{kazakov1993induced} is a model
defined on the $D$-dimensional square lattice: 
\begin{equation}
S_{\rm KM} = N
\sum_{x}\Tr 
\left\{m_0^2\Phi(x)^2
- \sum_{\mu=1}^D \Phi(x)U_\mu(x) \Phi(x+\mu) U^\dag_\mu(x)
\right\},
\label{eq:KMorg}
\end{equation}
where $\Phi(x)$ is a scalar field on the site $x$ and $U_\mu(x)$ is the link variable connecting
from $x$ to $x+\mu$. 
By integrating out the scalar fields, the effective theory becomes a summation of all possible Wilson loops.
Therefore this model is expected to induce QCD in an arbitrary dimension.

In this paper, we would like to generalize the square lattice on which the KM model is defined
to an arbitrary discretized space-time, which is described by a graph in general. 
Let us consider a directed graph $G$ and
suppose that a Hermitian matrix $\Phi_v$ and a unitary matrix $U_e \in U(N)$ live
on each vertex $v\in V$ and each edge $e\in E$, respectively.
Then we can naturally generalize the action of the KM model \eqref{eq:KMorg} to
\begin{equation}
S_{\rm gKM} = \beta
\Tr \left\{
\frac{1}{2}\sum_{v\in V}m_v^2\Phi_v^2
- q\sum_{e \in E}\Phi_{s(e)}U_e \Phi_{t(e)}U_e^\dag
\right\},
\label{KM action on the graph}
\end{equation}
where $\beta$, $m_v^2$ and $q$ are parameters of the model. 
We call this model as the graph KM (gKM) model in the following. 
Note that we can recover the original KM model by choosing the square lattice as the graph structure
and setting the parameters as $\beta=N$, $m_v^2=2 m_0^2$ (common for all the vertices) and $q=1$.
 
Apart from the induced QCD, 
this model can also be regarded as a model of a scalar field coupled to the gauge field on a discretized space-time (graph). 
To see this, we define the matrices, 
\be
L^q_U = T - q S_U, \qquad  \tilde{L}^q_U = S - q T_U,
\ee
where $S_U$ and $T_U$ are defined by (\ref{S and T}) with $X_e=U_e\otimes U_e^\dagger$, 
which are regarded as $q$-deformed covariant incidence matrices
associated with  forward and backward
difference operator on the graph, respectively.
Indeed, if we take $q=1$ and $N=1$ ($U_e=1$), $L^q_U$
is nothing but the incidence matrix,
and if we take $q=1$ only $L^q_U$ reduces to the covariant incidence
matrix discussed in \cite{Ohta:2021tmk}.
In \cite{Ohta:2021tmk}, it is also shown that
the covariant incidence matrix plays a role of the first order difference
operator and closely related to the Dirac operator on the graph, 
which reproduces the index theorem on the graph.

Using the $q$-deformed covariant incidence matrices, 
the action \eqref{KM action on the graph} can be written by
\be
S_{\rm gKM} = 
\beta
\Tr \left\{
\frac{1}{2}\sum_{v\in V} {m'_v}^2\Phi_v^2
- \frac{q}{2}\sum_{v,v' \in V}\Phi_{v} (\Delta^q_U)_{vv'}\Phi_{v'}
\right\},
\label{action and Laplacian}
\ee
where
\be
\begin{split}
\Delta^q_U  &= \frac{1}{2}\left(
L_U^{q\dag} L^q_U + \tilde{L}_U^{q\dag} \tilde{L}^q_U\right),\\
{m_v'}^2 &= m_v^2-\frac{1+q^2}{2}\deg v.
\label{q-deformed Laplacian and modified mass}
\end{split}
\ee
$\Delta^q_U$ without $U_e$ is called $q$-deformed graph Laplacian,
which becomes a square of the incidence matrix in the $q\to 1$ limit.
The graph Laplacian is a second order difference operator, which
is a discrete analog to the Laplace differential
operator on the continuum space-time.
As discussed in \cite{kazakov1993induced}, 
the action \eqref{action and Laplacian} would correspond to
the gauged scalar field theory without an explicit Yang-Mills term on a curved space-time; 
\be
S_{\rm gKM} \sim
\frac{N}{2g_0^2}\int d^D x \sqrt{g}
\Tr \Bigl\{
R(x)\Phi(x)^2
- \Phi(x) \Delta_U\Phi(x)
\Bigr\},
\ee
where $\Delta_U\equiv D_\mu D^\mu$ is the gauge covariant
Laplace operator and $R(x)$ is a scalar function
depending on the position $x$, in a suitable continuum and $q\to 1$ limit.



\begin{figure}[H]
\begin{center}
\subcaptionbox{Triangular lattice}[.3\textwidth]{
\includegraphics[scale=0.4]{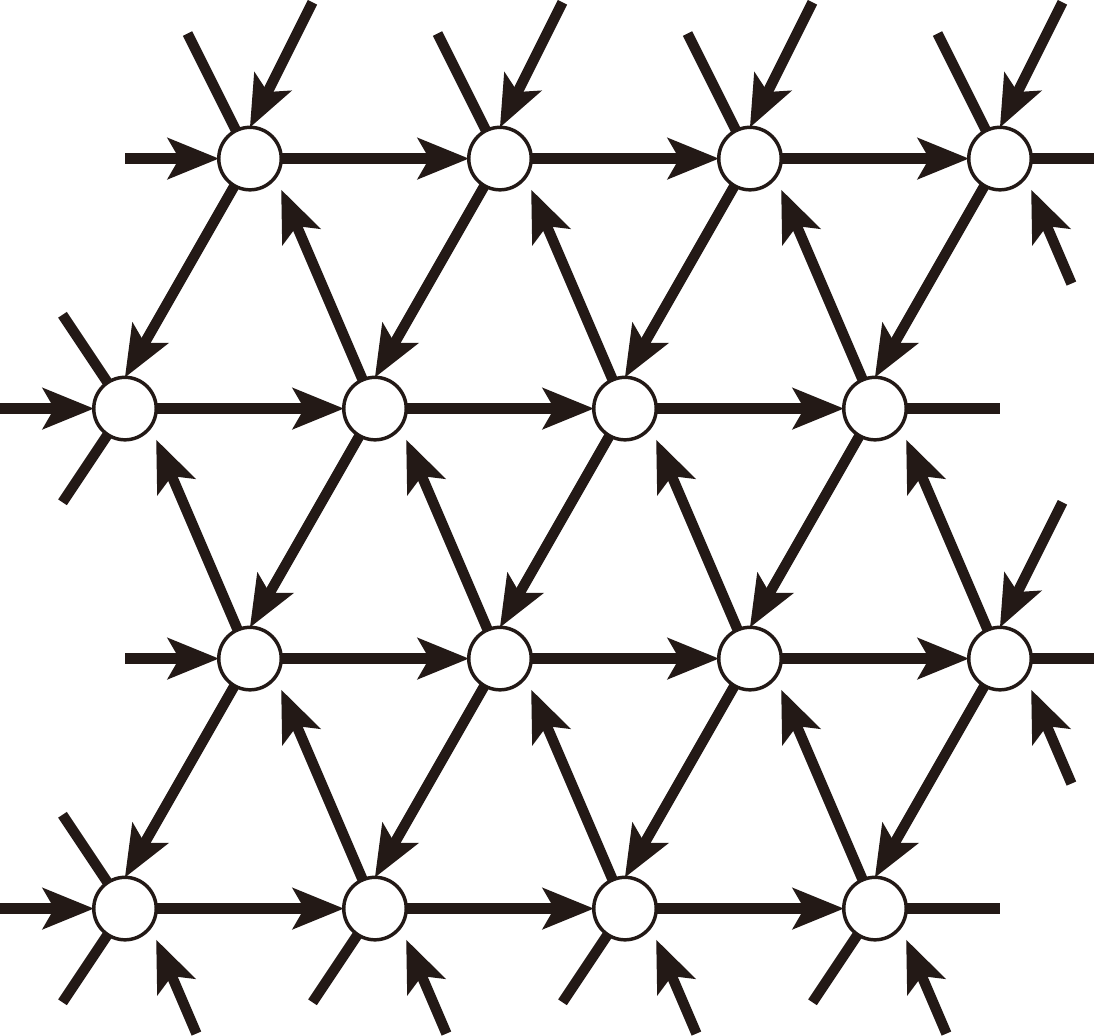}
}
\subcaptionbox{Square lattice}[.3\textwidth]{
\includegraphics[scale=0.4]{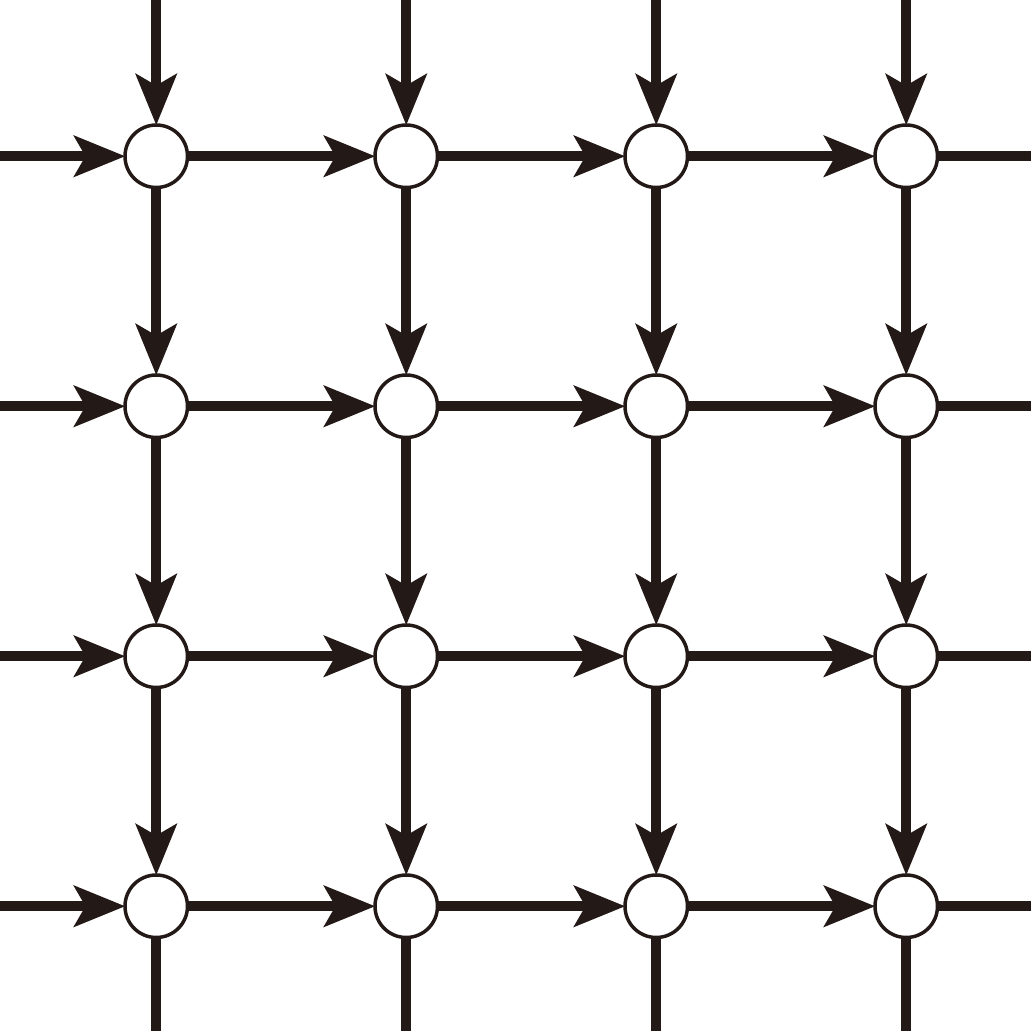}
}
\subcaptionbox{Hexagonal lattice}[.3\textwidth]{
\includegraphics[scale=0.3]{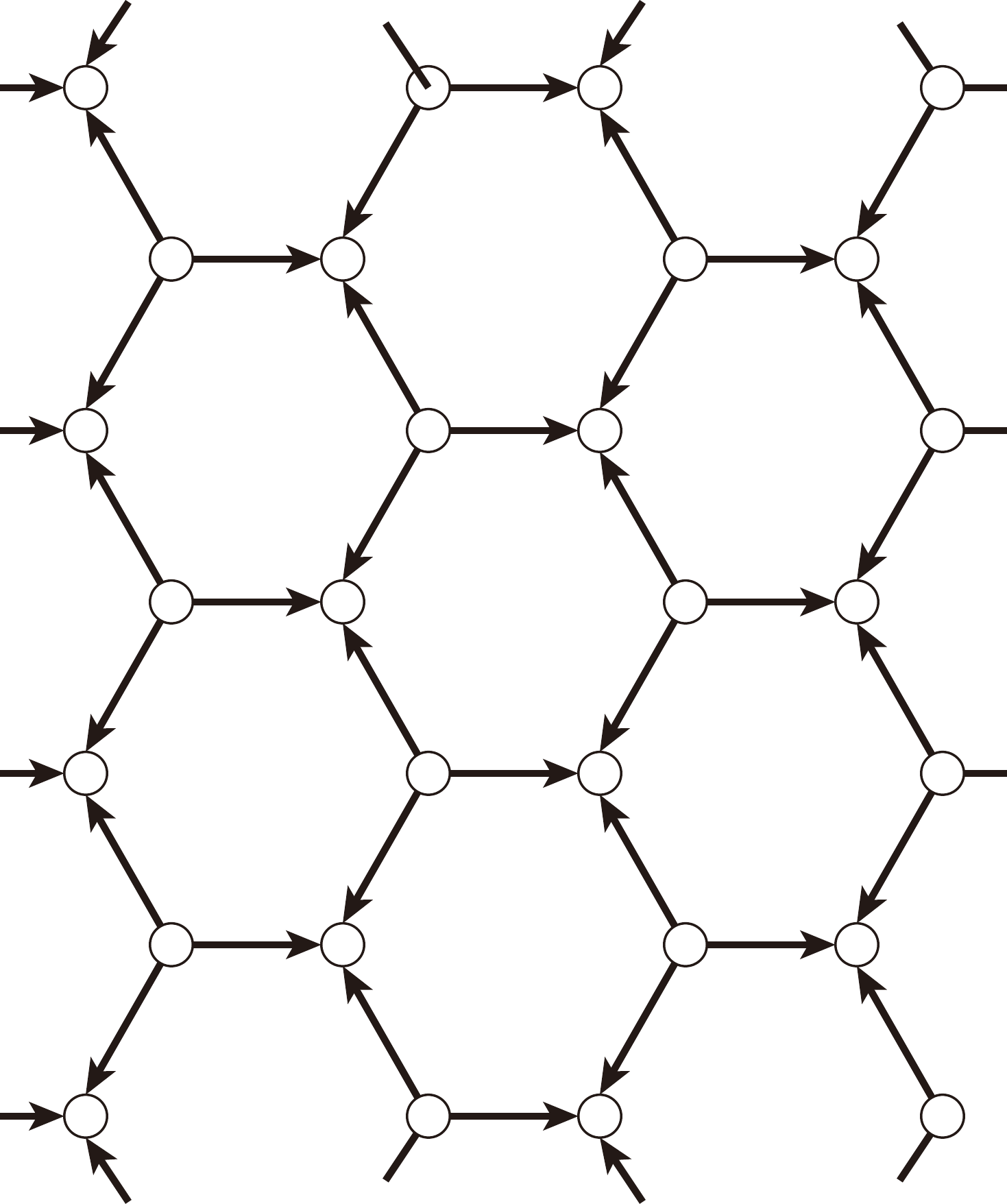}
}
\end{center}
\caption{Discretizations of a flat plane or torus $T^2$ can be obtained from graphs.
The original KM model is defined on the square lattice graph (b).
In this sence, the gKM model has possibilities
for constructing various kind of discretized scalar field theories by choosing the graph
structure.}
\label{2D discretization}
\end{figure}

The advantage of being able to define a theory on an arbitrary graph is not only the ability to discretize a curved space-time. Even in considering the flat space-time, the advantage of having a various kind of discretizations is significant. For example, the flat space-time in two dimensions can be discretized by graphs with a periodic structure as shown in Fig.~\ref{2D discretization}. Of course, the original KM model is included in these, but this extension allows the theory to have greater discrete symmetry.
This will be also useful when considering applications to condensed matter physics in particular.




\subsection{Parameter tuning}

Now let us discuss the relationship between the graph Kazakon-Migdal model and the Ihara zeta function on the graph. 
First of all, since the integrand is Gaussian with respect to $\Phi_v$, we can integrate over $\Phi_v$ of the partition function
\begin{equation}
Z_{\rm gKM} = \int \prod_{v\in V}d\Phi_v \prod_{e\in E} dU_e \,
\, e^{-S_{\rm gKM}},
\label{eq:partition function}
\end{equation}
then it reduces explicitly to
\begin{equation}
Z_{\rm gKM} = \left(\frac{\beta}{2\pi}\right)^{-\frac{n_V N^2}{2}}\int \prod_{e\in E} dU_e\,
{\det}\lrpar{M-q A_U}^{-1/2}\,,
\label{eq:Zmed}
\end{equation}
where $M$ is a diagonal matrix given by $M=m_v^2\delta_{vv'}\otimes \bs{1}_{N^2}$
and $A_U$ is the extended adjacency matrix given by \eqref{eq:adjacency matrix X}
with $X_e=U_e\otimes U_e^\dagger$.

Here, motivated from the expression \eqref{eq:extended Ihara2}, if we set
\begin{equation}
  m^2_v = 1+ \left(\deg v-1\right) q^2 \,,
  \label{eq:para1}
\end{equation}
then the partition function \eqref{eq:Zmed} can be expressed through the
extended Ihara zeta function;
\begin{align}
  Z_{\rm gKM} &= \left(\frac{\beta}{2\pi}\right)^{-\frac{n_V N^2}{2}}\int \prod_{e\in E} dU_e\,
{\det}\lrpar{1_{n_V}\otimes 1_{N^2}-q A_U + q^2(D-1_{n_V})}^{-\frac{1}{2}} \nn \\
  &= \left(\frac{\beta}{2\pi}\right)^{-\frac{n_V N^2}{2}}
  (1-q^2)^{\frac{(n_E-n_V)N^2}{2}}\int \prod_{e\in E} dU_e\, \zeta_G(q;U)^\frac{1}{2}\,.
  \nn
\end{align}
We can eliminate the $q$-dependence in the overall factor by further setting
\begin{align}
  \beta = 2\pi\alpha (1-q^2)^{\frac{n_E}{n_V}-1} \,,
  \label{eq:para2}
\end{align}
with a $q$-independent parameter $\alpha$.
Note that if we use the $q$-deformed graph Laplacian, the modified mass
in \eqref{q-deformed Laplacian and modified mass} becomes
\be
{m_v'}^2 = \frac{1}{2}(q^2-1)\left(\deg v -2\right),
\ee
which represents massless field when $\deg v =2$ for all $v$ or $q=1$.

Since the extended Ihara zeta function can be expressed as \eqref{eq:extended Ihara P},
we see that the partition function of the gKM model
\eqref{eq:partition function} with the parametrization \eqref{eq:para1} and \eqref{eq:para2}
becomes
\begin{align}
  Z_{\rm gKM} &= \int \prod_{v\in V}d\Phi_v \prod_{e\in E} dU_e
  \, e^{-2\pi\alpha (1-q^2)^{\frac{n_E}{n_V}-1}
  \Tr \Bigl(
  \frac{1}{2}\sum_{v\in V}(1+q^2(\deg v-1))\Phi_v^2
  - q\sum_{e \in E}\Phi_{s(e)}U_e \Phi_{t(e)}U_e^\dag
  \Bigr)} \nn \\
  &=
  \alpha^{-\frac{n_V N^2}{2}} \int \prod_{e\in E} dU_e\,
  e^{ \frac{1}{2}\sum_{k=1}^\infty \frac{q^k}{k}\sum_{C\in R_k}{\Tr P_{C}(U)\, \Tr P_{C}(U)^\dagger}
    }\nn \\
    &\equiv \alpha^{-\frac{n_V N^2}{2}} Z_G(q) \,. 
    \label{eq:Zq}
\end{align}
Since $\Tr P_C(U)$ is the Wilson loop along a reduced cycle $C$,
this matrix integral is regarded as the generating functional of the integral of the Wilson loops on the graph of constant length\footnote{
This kind of the unitary matrix integral also appears
in the context of a generalization of the two-dimensional lattice gauge theory
\cite{Gross:1980he,Wadia:1980cp}
or the thermodynamics of (supersymmetric) gauge theory
\cite{Hallin:1998km,Sundborg:1999ue,Dutta:2007ws}.
}.
We can of course express $Z(q)$ by using the extended Ihara zeta function; 
\[
  Z_G(q) = \int \prod_{e\in E} dU_e\, \zeta_G(q;U)^\frac{1}{2}\,.
\]

Note that the specific parameter settings \eqref{eq:para1} and \eqref{eq:para2} are essential for the partition function of the gKM model to count the Wilson loops without over- or under-counting.
As an example, 
let us consider the case when $\beta={\rm const.}$ and $m_v^2$ is a constant independent of $v$, as in the original KM model.
In this case, the partition function becomes
\begin{align}
  Z_{\rm gKM} &\sim \int \prod_{e\in E} dU_e\,
  e^{\frac{1}{2}\sum_{k=1}^\infty \frac{q^k}{k} \Tr A_U^k  }\,.
  \nn
\end{align}
The point is that $\Tr A_U^k$ includes not only the Wilson loops of length $n$
but also some extra constant coming from the backtrackings.
A typical example is the triangle graph whose (extended) adjacency matrix is given by
\[
  A_U=\begin{pmatrix}
  0                       & U_1\otimes U_1^\dagger  & U_3^\dagger \otimes U_3 \\
  U_1^\dagger \otimes U_1 & 0                       & U_2\otimes U_2^\dagger \\
  U_3 \otimes U_3^\dagger & U_2^\dagger \otimes U_2 & 0
\end{pmatrix}\,.
\]
Although it is easy to see $\Tr A_U^2 = 6 N^2$,
it is obvious that the triangle graph does not have any Wilson loop of length 2; 
it does not count Wilson loops but rather counts the backtrackings. 
Of course, if we impose a general $q$-dependence on $m_v^2$ or $\beta$, the relationship between the power of $q$ and the length of the Wilson loop would be more complicated.
This suggests that the parameter settings \eqref{eq:para1} and \eqref{eq:para2} are crucial for
the partition function of the gKM model \eqref{eq:partition function}
counts the Wilson loops correctly.

\subsection{Duality between unitary and Hermitian matrix integrals}

We have obtained the expression \eqref{eq:Zmed}
(or \eqref{eq:Zq} after the parameter settings)
by integrating the Hermitian matrices $\Phi_v$ first.
Instead, we can express the partition function \eqref{eq:partition function}
by integrating the unitary matrices $U_e$ first \cite{kazakov1993induced}
by using the HCIZ integral formula \cite{itzykson1980planar},
\begin{equation}
  \int dU \, e^{t \Tr AUBU^\dagger}
  = \frac{G(N+1)}{t^\frac{N^2-N}{2}}
  \frac{{\det}_{i,j} \lrpar{e^{t a_i b_j}}} {\Delta(a)\Delta(b)}\,,
  \label{eq:IZ}
\end{equation}
where
$U$ is a unitary matrix of size $N$, 
$dU$ is the Haar measure of $U(N)$ normalized as $\int dU = 1$, 
$G(N+1)=\prod_{i=1}^{N-1} i!$ is the Barnes function,
$A$ and $B$ are Hermitian matrices whose eigenvalues are $(a_1,\cdots,a_N)$
and $(b_1,\cdots,b_N)$, respectively,
and $\Delta(a)$ and $\Delta(b)$ are the Vandermonde determinant with respect to $A$ and $B$,
respectively;
\[
  \Delta(a) = \prod_{1\le i<j\le N} (a_j-a_i)\,, \quad
  \Delta(b) = \prod_{1\le i<j\le N} (b_j-b_i)\,.
\]

We apply the formula \eqref{eq:IZ} to the partition function of the gKM
model \eqref{eq:partition function}.
Since the integrand includes only the eigenvalues of
the Hermitian matrices $\Phi_v$, we also use a mapping
from the matrix integral to the
integral over the eigenvalues 
(see e.g.\cite{marino2004houches}),
\[
  \int d\Phi f(\Phi) =
  \frac{(2\pi)^{\frac{N(N-1)}{2}}}{G(N+2)} \int \prod_{i=1}^N d\phi_i\,
  \Delta(\phi)^2 f(\phi)\,,
\]
for a Hermitian matrix $\Phi$ with eigenvalues $\phi_i$ ($i=1,\cdots,N$).
We then obtain
\begin{multline}
  Z_{\rm gKM}
  =
  \frac{{G(N+1)}^{n_E-n_V}} {(N!)^{n_V}}
  \frac{(2\pi)^{\frac{n_V N(N-1)}{2}}}
       {\beta^{\frac{N^2}{2}n_V} q^{\frac{n_E N(N-1)}{2}}}
  \int \lrpar{ \prod_{v\in V}\prod_{i=1}^N d\phi_{v,i} }
  \lrpar{\prod_{v'\in V}\Delta(\phi_{v'})^{2-\deg v'}} \\
  \times \exp\lrpar{ {-\frac{1}{2}\sum_{v''\in V}\sum_{j=1}^N m_{v''}^2 \phi_{v'',j}^2} }
  \prod_{e\in E} \underset{i,j}{\det}\lrpar{e^{q \phi_{s(e),i}\phi_{t(e),j}}}\,,
  \nn
\end{multline}
where
$\phi_{v,i}$ ($v\in V,\ i=1,\cdots,N$) are the eigenvalues of $\Phi_v$
rescaled so that the parameter $\beta$ is included only in the overall factor.
In particular, after setting \eqref{eq:para1} and \eqref{eq:para2},
the partition becomes as \eqref{eq:Zq}
and $Z_G(q)$ can be expressed in two ways as
\begin{align}
\label{eq:dual1}
  Z_G(q) &= \int \prod_{e\in E} dU_e\,
  e^{ \frac{1}{2}\sum_{k=1}^\infty \frac{q^k}{k}\sum_{C\in R_k}{\Tr P_{C}(U)\, \Tr P_{C}(U)^\dagger}} \\
  &=
  \frac{G(N+1)^{n_E-n_V}}
       {(2\pi)^\frac{n_V N}{2} (N!)^{n_V} q^\frac{n_E N(N-1)}{2} (1-q^2)^{\frac{(n_E-n_V)N^2}{2}}}
  \int \lrpar{ \prod_{v\in V}\prod_{i=1}^N d\phi_{v,i} }
  \lrpar{\prod_{v'\in V}\Delta(\phi_{v'})^{2-\deg v'}} \nn \\
  &\hspace{2.5cm}\times \exp\lrpar{ {-\frac{1}{2}\sum_{v''\in V}
  \sum_{j=1}^N \lrpar{1+(\deg v''-1)q^2} \phi_{v'',j}^2} }
  \prod_{e\in E} \underset{i,j}{\det}\lrpar{e^{q \phi_{s(e),i}\phi_{t(e),j}}}\,.
  \label{eq:dual2}
\end{align}
In the next section, we evaluate this integral exactly in the large $N$ limit.
Before that, however, we consider special cases when we can evaluate this integral at finite $N$.

\subsection{The dual integral in cycle graphs}
\label{sec:ZC1}

\begin{figure}[H]
\begin{center}
\subcaptionbox{$C_n$ ($n$-gon)}[.45\textwidth]{
\includegraphics[scale=0.4]{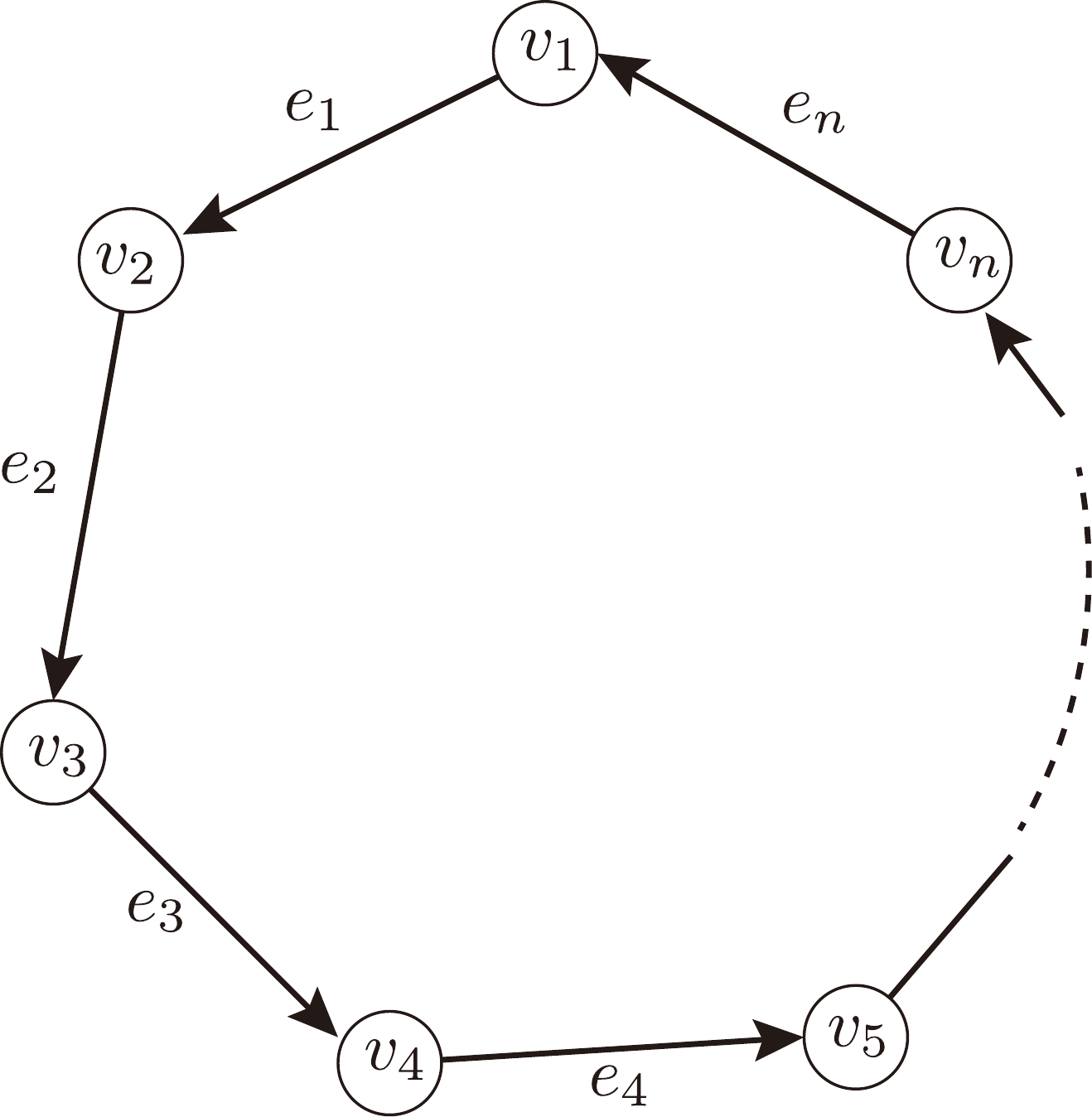}
}
\subcaptionbox{$C_1$ (1-gon)}[.45\textwidth]{
\includegraphics[scale=0.5]{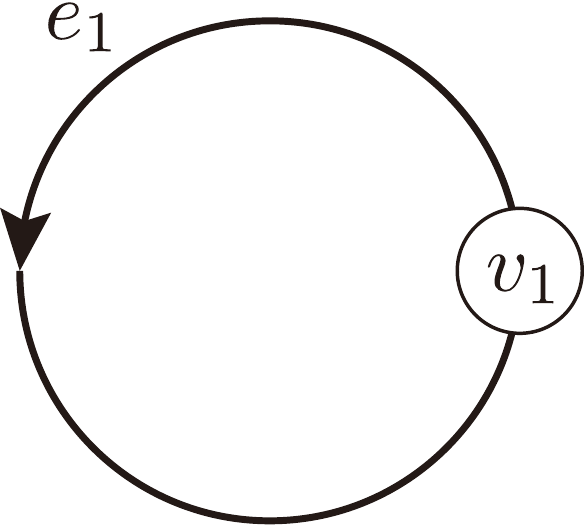}
}
\end{center}
\caption{A cycle graph $C_n$ has $n$ vertices and edges. This is also called $\hat{A}_{n-1}$ quiver chain.
(b) is a simplest case of $C_1$ (1-gon). $Z_{C_n}$ reduces to $Z_{C_1}$ with the length $n$ cycle.}
\label{Figure of the cycle graph}
\end{figure}

The integrals \eqref{eq:dual1} and \eqref{eq:dual2} are difficult to carry out in general,
but there are exceptions; it becomes a Gaussian integral when $\deg v=2$
for all the vertices, which is achieved when $G$ is a cycle graph (a single polygon).
Let us consider the case when $G$ is a cycle graph $C_n$ ($n$-gon).
(See Fig.~\ref{Figure of the cycle graph}.)
In this case, the number of the vertices and edges are the same, $n_V=n_E=n$,
and we have the same number of unitary matrices $U_e$ and Hermitian matrices $\Phi_v$
$(e,v=1,\cdots,n)$.

We first evaluate the integral \eqref{eq:dual1}.
Since the cycle graph $C_n$ has only one chiral primitive cycle $e_1 \cdots e_n$ as a representative of the equivalence class, the length of a general reduced cycle is a multiple of $n$, and there are $2n$ reduced cycles of length $nk$ in total, that is, $k$-power of $n$ chiral primitive cycles and their inverses.
Therefore, the integral \eqref{eq:dual1} becomes
\begin{align}
  Z_{C_n}(q) &= \int \prod_{e=1}^n dU_e\,
    e^{\frac{1}{2}\sum_{m=1}^\infty \frac{q^m}{m}\sum_{C\in R_m}
    {\Tr P_{C}(U)\, \Tr P_{C}(U)^\dagger}}  \nn \\
  &=
  \int \prod_{e=1}^n dU_e\,
  e^{ \frac{1}{2}\sum_{k=1}^\infty \frac{q^{nk}}{nk}
  \times 2n\Tr (U_1\cdots U_n)^k \Tr(U_n^\dagger \cdots U_1^\dagger)^k} \nn \\
  &=
  \int dU\,
  e^{ \sum_{k=1}^\infty \frac{(q^{n})^k}{k}
  \Tr U^k \Tr U^\dagger{}^k } \nn \\
  &= Z_{C_1}(q^n)\,,
  \label{eq:ZCn}
\end{align}
where the integration over the $n$ unitary matrices has been
reduced to an integration over a single unitary matrix $U$
in the last line by using the left-right invariance of the Haar measure $dU$ and $\int dU=1$.
Note that the similar type of integral, which has more
generic coefficients in the exponent,
is evaluated by the Frobenius formula in \cite{Dutta:2007ws}. It is remarkable that we will obtain the more explicit result despite finite $N$
thanks to the dual description.

We can determine the function $Z_{C_1}(q)$ at finite $N$ by evaluating
the integral \eqref{eq:dual2} for $n=1$ (1-gon);
%
%
\begin{align}
  Z_{C_1}(q)
  &= \frac{1}{(2\pi)^\frac{N}{2} N!\, q^{\frac{N(N-1)}{2}}}
  \sum_{\sigma \in S_N} \sgn(\sigma)
   \int \prod_{i=1}^N d\phi_i\, 
   e^{-\frac{1}{2} \sum_{i,j=1}^N\phi_{i}\phi_{j}
    \lrpar{
    (1+q^2)\delta_{ij} -q \delta_{i,\sigma(j)}-q \delta_{j,\sigma(i)}
    }} \nn \\
  &= \frac{1}{ N!\, q^{\frac{N(N-1)}{2}}}
  \sum_{\sigma \in S_N} \sgn(\sigma)
  \left(
  \det_{i,j}\left( (1+q^2)\delta_{ij} -q \delta_{i,\sigma(j)}-q \delta_{j,\sigma(i)} \right)
  \right)^{-1/2} \nn \\
  & \equiv \frac{1}{ N!\, q^{\frac{N(N-1)}{2}}}
  \sum_{\sigma\in S_N} \sgn(\sigma) \left( \det M_\sigma \right)^{-1/2}\,,
  \label{eq:I1_med}
\end{align}
where $M_\sigma$ is a matrix of size $N$.

The matrix with the $i$-th and $j$-th rows and columns of $M_\sigma$ simultaneously replaced is $M_{\tau_{ij}\sigma\tau_{ij}}$, where $\tau_{ij}\in S_N$ is the transposition of $i$ and $j$,
which does not change the determinant.
Therefore, assuming that $\sigma$ is an element of the equivalent class labeled by the partition $\lambda=(\lambda_1,\cdots,\lambda_{r'})$,
we can transform $M_\sigma$ to the following standard form without changing the determinant
by acting an appropriate permutation:
\[
  M_\sigma \to M_\lambda \equiv \begin{pmatrix}
  K_{\lambda_1}  \\
  & \ddots & \\
  && K_{\lambda_{r'}}
\end{pmatrix}\,,
\]
where $K_m$ ($m\in \N$) are matrices with size $m$ given by
\[
  K_1 \equiv (q-1)^2, \quad
  K_2 \equiv \begin{pmatrix} q^2+1 & -2q \\ -2q & q^2+1 \end{pmatrix}, \quad
  K_m \equiv \begin{pmatrix}
  q^2+1  & -q    & \cdots & 0 & -q  \\
  -q     & q^2+1 & \cdots & 0 & 0 \\
  \vdots & \vdots& \ddots & \vdots & \vdots \\
  0     & 0      & \cdots & q^2+1   & -q \\
  -q     & 0     & \cdots & -q     & q^2+1
\end{pmatrix}\ (m\ge 3)\,.
\]
Since it is straightforward to show
\[
  \det K_m = (1-q^m)^2\,, \quad (m\in\N)
\]
we see
\[
  \det M_\lambda = \prod_{i=1}^{r'} (1-q^{\lambda_i})^2\,.
\]
Since the summation $\sum_{k=1}^\infty \frac{q^k}{k}|\Tr U^k|^2$ 
in \eqref{eq:ZCn} converges only for $|q|<1$,
\eqref{eq:I1_med} is evaluated as
\begin{align}
  Z_{C_1}(q) &=
  \frac{1}{N!\,q^{N(N-1)/2}}
  \sum_{\lambda \vdash N} \sgn(\lambda) n_\lambda \prod_{i=1}^{r'} \frac{1}{|1-q^{\lambda_i}|} \nn \\
  &=
  \frac{1}{q^{N(N-1)/2}}
  \sum_{\lambda \vdash N} \frac{n_\lambda}{N!} \prod_{i=1}^{r'} \frac{-(-1)^{\lambda_i}}{|1-q^{\lambda_i}|} \nn \\
  &=
  \displaystyle \frac{(-1)^N}{q^{N(N-1)/2}}
  \sum_{\lambda \vdash N} \frac{n_\lambda}{N!} \prod_{i=1}^{r'} \frac{1}{q^{\lambda_i}-1}\,, 
\label{eq:ZC1 result}
\end{align}
where 
$\sgn(\lambda)=\prod_{i=1}^{r'}(-1)^{\lambda_i-1}$ and
$n_\lambda$ 
is  the number of conjugates of a permutation $\sigma\in S_N$
labeled by the partition (Young diagram) $\lambda$.
The partition $\lambda$ is also given by a set of the length $l_i$ and the multiplicity $m_i$ of the rows of the Young diagram as $\lambda=(l_1^{m_1}\cdots l_r^{m_r})$. 
Using this notation, if we define
\begin{equation}
  z_\lambda = \prod_{i=1}^r {m_i!\, l_i^{m_i}}\,,
  \label{eq:zlambda}
\end{equation}
$n_\lambda$ is expressed by
$\frac{N!}{z_\lambda}$.
Thus, applying the formula proved in Appendix \ref{app:proof};
\begin{equation}
  \sum_{\lambda\vdash N} \frac{1}{z_\lambda}\prod_{i=1}^{r'}\frac{1}{q^{\lambda_i}-1 }
   = {q^{N(N-1)/2}}\prod_{i=1}^N \frac{1}{q^i-1}\,, 
  \label{eq:th1-2}
\end{equation}
we finally obtain
\begin{align}
  Z_{C_1}(q) = \int dU \, e^{\sum_{k=1}^\infty \frac{q^k}{k}\Tr U^k \Tr U^{\dagger k}}
&=
  \displaystyle   \prod_{i=1}^N \frac{1}{1-q^i}\,, 
\label{eq:ZC1 result0}
\end{align}
for any finite $N$ and $|q|<1$.

We make a comment on the region of $q$. 
Although the expression \eqref{eq:ZCn} is well defined only in the region 
$|q|<1$, the dual expression \eqref{eq:I1_med} is defined also in $|q|>1$.
Using the formula proved again in Appendix \ref{app:proof},
\begin{equation}
  \sum_{\lambda\vdash N} \frac{1}{z_\lambda}\prod_{i=1}^{r'}\frac{1}{1-q^{\lambda_i} } = \prod_{i=1}^N \frac{1}{1-q^i}\,, 
  \label{eq:th1}
\end{equation}
we can evaluate \eqref{eq:I1_med} in the region $|q|>1$ as
\begin{align}
  Z_{C_1}(q) = \prod_{i=1}^N \frac{1}{q^{i-1}-q^{2i-1}}\,.
  \label{eq:ZC1 result qlt1}
\end{align}
In Appendix \ref{app:q1}, we evaluate the unitary matrix integral
\eqref{eq:ZCn} directly by using Cauchy's integral formula and reproduce 
the result \eqref{eq:ZC1 result0} for $|q|<1$. 
Interestingly enough, in the same appendix, we also evaluate 
the integral for $|q|>1$ ignoring the convergence, 
which reproduces \eqref{eq:ZC1 result qlt1}. 
We thus expect that the unitary matrix integral \eqref{eq:ZCn} can be 
analytically connected to the region $|q|>1$.
We will also discuss this point in the last section.

\section{The matrix integral in the large $N$ limit}
\label{sec:large N}

In this section, we evaluate $Z_G(q)$ in the large $N$ limit.
 In preparation for this, 
 we start with showing some properties satisfied by unitary matrix integrals
 at large $N$.

To simplify the description, we introduce the notation,
\[
  |\Tr f(U) |^2 \equiv \Tr f(U) \,\Tr f(U)^{\dagger}\,,
\]
and
\[
  \Bigl\langle
   g(U)
  \Bigr\rangle \equiv \int \prod_{e\in E} dU_e\, g(U)\,,
\]
where $f(U)=f(U_1,\cdots,U_{n_E})$ and 
$g(U)=g(U_1,\cdots,U_{n_E})$ are functions of $U_e$ ($e=1,\cdots,n_E$).

\subsection{Integrals of Wilson loops in the large $N$ limit}
\label{sec:decomp}
\subsubsection*{\underline{\it Fundamental integral and its behavior at large $N$}}

The most fundamental integral in performing general unitary matrix integrals is
\begin{equation}
  \int dU U_{i_1j_1}\cdots U_{i_nj_n} U^*_{i'_1j'_1}\cdots U^*_{i'_mj'_m}
  = \delta_{m,n} \sum_{\sigma,\tau\in S_n}
  \delta_{i_1,i'_{\sigma(1)}}\cdots \delta_{i_n,i'_{\sigma(n)}}
  \delta_{j_1,j'_{\tau(1)}}\cdots \delta_{j_n,j'_{\tau(n)}}
  W\!g(\sigma\tau^{-1},N)\,,
  \label{eq:basic int}
\end{equation}
where $W\!g(\sigma,N)$ is the Weingarten function \cite{weingarten1978asymptotic,collins2003moments},
\[
  W\!g(\sigma,N) = \frac{1}{n!^2}\sum_{\lambda\vdash n}
  \frac{\chi^\lambda(1)^2\chi^\lambda(\sigma)}{s_{\lambda,N}(1)}\,,
\]
where $\chi^\lambda(\sigma)$ is the character of $S_n$
corresponding to the partition $\lambda$
and $s_{\lambda,N}$ is the Schur polynomial of $\lambda$. 
Since we are interested in the large $N$ limit,
we do not need the detail of this function
but need only its asymptotic behavior in $N\to\infty$,
which is given by \cite{collins2006integration},
\begin{align}
  W\!g(\sigma,N) = \frac{1}{N^{n+|\sigma|}} \prod_{i=1}^k (-1)^{\ell(c_i)-1}{\rm Cat}(\ell(c_i)-1) +{\cal O}(N^{-n-|\sigma|-2})\,,
  \label{eq:WGasymp}
\end{align}
where we have assumed $\sigma\in S_n$ is a product of cycles $\sigma=c_1\cdots c_k$,
$|\sigma|$ denotes the minimal number of factors necessary to write $\sigma$
as a product of transpositions,
and ${\rm Cat}(m)=\frac{(2m)!}{m!(m+1)!}$ is the Catalan number.
From \eqref{eq:WGasymp}, we see that the leading contributions
of the $1/N$ expansion of the integral \eqref{eq:basic int}
come from $\tau=\sigma$ and thus it can be written as
\begin{multline}
  \qquad
  \int dU \, U_{i_1j_1}\cdots U_{i_nj_n} U^*_{i'_1j'_1}\cdots U^*_{i'_nj'_n}\\
  = \frac{1}{N^n} \sum_{\sigma\in S_n}
  \delta_{i_1,i'_{\sigma(1)}}\cdots \delta_{i_n,i'_{\sigma(n)}}
  \delta_{j_1,j'_{\sigma(1)}}\cdots \delta_{j_n,j'_{\sigma(n)}}
  + {\cal O}(N^{-n-1})\,.
  \qquad
  \label{eq:basic asymp}
\end{multline}
Note that this is also obtained in \cite{Kazakov:1983fn, KOSTOV1984191, OBRIEN1985621}. 
All the properties we show in this subsection follow from this expression.

\subsubsection*{\underline{\it Integral of a single Wilson loop}}

Let us consider a Wilson loop.
Recalling that a Wilson loop is associated with an equivalent class of reduced cycles and a reduced cycle is consist of a power of a primitive cycle,
the Wilson loop is written as $\Tr P_C(U)^l$ for a primitive loop $C$ in $G$ and an integer $l$.
We can then evaluate the integral of $|\Tr P_C(U)^{l}|^2$
at large $N$ as
\begin{equation}
  \Bigl\langle 
    |\Tr P_C(U)^{l} |^2
  \Bigr\rangle \underset{N\to\infty} \longrightarrow l \,. 
  \label{eq:PCl}
\end{equation}

To show it, we pick one of the unitary matrices
included in $P_C(U)$ and call it $U$.
Then $\Tr P_C(U)$ can be written in general as
\begin{align}
  \Tr P_C(U) = \Tr\left(
  U^{\epsilon_1} A_1 U^{\epsilon_2}A_2 \cdots U^{\epsilon_r} A_r\right)\,,
  \label{eq:UinP}
\end{align}
where $\epsilon_i=\pm 1$ ($i=1,\cdots,r$) and $A_i$ is a product of $U_e$ $(e\in E)$ except for $U$.
Since $C$ is a primitive cycle, all $A_i$ are never the same.
In the following, we assume $\epsilon_i=1$ for all of $i=1,\cdots,r$
because, as we will see soon, the signatures are irrelevant in the following discussion.

If there is such a unitary matrix $U$ that satisfies $r=1$,
the left-hand side of \eqref{eq:PCl} is reduced
to an integral over a single unitary matrix
by absorbing all the other unitary matrices
in $P_C(U)$ to $U$ via the left-right invariance of the Haar measure.
Then \eqref{eq:PCl} immediately follows from
the formula given in \cite{diaconis1994eigenvalues,diaconis2001linear},
\begin{equation}
  \Bigl\langle
    \bigl|\bigl(\Tr U^{l_1}\bigr)^{m_1} \cdots
    \bigl(\Tr U^{l_r}\bigr)^{m_r}\bigr|^2
  \Bigr\rangle = z_\lambda,
  \quad (N\ge |\lambda|)\,,
  \label{eq:DS formula}
\end{equation}
where $\lambda=(l_1^{m_1}\cdots l_r^{m_r})$ is
a partition,
$|\lambda|=\sum_{i=1}^rl_im_i$,
and $z_\lambda$ is defined by \eqref{eq:zlambda}.
However, $r$ is not always $1$ in general
and so we have to evaluate the integral more carefully.

Let us first show the case of $l=1$, which includes all the essence.
From \eqref{eq:basic asymp}, the left-hand side of \eqref{eq:PCl} is written as
\begin{align}
\Bigl\langle 
\bigl|\Tr P_C(U)\bigr|^2
\Bigr\rangle
  = \int \prod_{\underset{U\ne U_e}{e\in E}} dU_e
   \frac{1}{N^r}\Bigl\{  \sum_{\sigma\in S_r}
   \Bigl(
  &\delta_{i_1,i'_{\sigma(1)}}\cdots \delta_{i_n,i'_{\sigma(r)}}
  \delta_{j_1,j'_{\sigma(1)}}\cdots \delta_{j_n,j'_{\sigma(r)}}\nn \\
  &\times (A_1)_{j_1i_2}\cdots(A_r)_{j_r i_1}
  (A^\dagger_r)_{i'_1j'_r}\cdots(A^\dagger_1)_{i'_2j'_1}\Bigr)
  + {\cal O}(1/N)\Bigr\}\,,
  \label{eq:expandPC}
\end{align}
where ${\cal O}(1/N)$ includes terms with different permutations among $i$'s and $j$'s.
For each $\sigma$, we obtain a product of traces of $A_i$ and $A_i^\dagger$ but
the maximal number of the traces in the product is $r$ from the construction.

The point is that the trace of $A_i$ and $A_i^\dagger$ gives $\Tr(A_iA_i^\dagger)=N$
because $A_i$'s are unitary matrices and this is the only chance to give an ${\cal O}(N)$ contribution.
Therefore the maximal power of $N$ in the summation of $\sigma\in S_n$ is
$\Tr(A_1A_1^\dagger)\cdots\Tr(A_rA_r^\dagger)=N^r$ and it occurs only once in the summation
because $C$ is now assumed to be a primitive cycle,
and the other contributions are at most of the order of $1/N$ compared to this term.
In fact, in the above expression, $\sigma=1$ gives the contribution $N^r$
and other terms consist of a product of traces times a power of $N$ less than $r$.
Note that, if we take the signatures $\epsilon_i$ into account,
the permutation $\sigma$ which gives $N^r$ is not $\sigma=1$ in general,
but the structure is the same: Only one element of $S_r$ gives the contribution $N^r$ and the other terms are ${\cal O}(1/N)$ compared to it.
This is the reason why the signatures $\epsilon_i$ are irrelevant in this discussion.
Therefore \eqref{eq:expandPC} becomes
\begin{align}
  \Bigl\langle 
    \bigl|\Tr P_C(U) \bigr|^2
  \Bigr\rangle
  = \int \prod_{\underset{U\ne U_e}{e\in E}} dU_e\,
  \Bigl(
    1 + {\cal O}(1/N)\times [\text{terms with traces of $A_i$ and $A_i^\dagger$}]
  \Bigr).
  \nn
\end{align}
The remaining ${\cal O}(1/N)$ terms include functions of other unitary matrices than $U$.
However, they disappear in the large $N$ limit.
This is because the integral of such a function is of the order of unity in general
and the number of terms appearing in this part is ${\cal O}(N^0)$,
since we are considering sufficiently large $N$ than $r$.
For the same reason, we can ignore the ${\cal O}(1/N)$ terms in \eqref{eq:expandPC}.
We can then conclude \eqref{eq:PCl} for $l=1$.

We next consider the case of $l>1$.
In this case, the leading contribution
$(\Tr(A_1A_1^\dagger)\cdots\Tr(A_rA_r^\dagger))^l=N^{rl}$
comes not only from $\sigma=1$ but also from
other $l-1$ permutations,
\[
  \sigma= \begin{pmatrix}
  D_1 & D_2 & \cdots & D_l \\
  D_i & D_{i+1} & \cdots & D_{i-1}
\end{pmatrix}\,, \quad (i=2,\cdots,l)
\]
where $D_i=((i-1)r+1,\cdots, ir)$ and the indices of $D$ are
assumed to be written in ${\rm mod}\ l$.
Therefore, as the same reason of the case of $l=1$,
we can conclude \eqref{eq:PCl} for $l>1$.

\subsubsection*{\underline{\it Large $N$ decomposition}}

We next consider two Wilson loops
$\Tr(P_{C_1}(U))$ and $\Tr(P_{C_2}(U))$ corresponding to
two (not necessarily primitive) cycles $C_1$ and $C_2$, respectively.
We can show that the integral of the product of
the two Wilson loops are decomposed into the product of
the integrals of the individual Wilson loops
at large $N$:
\begin{align}
  \Bigl\langle
    \bigl|\Tr P_{C_1}(U)
    \Tr P_{C_2}(U) \bigr|^2
  \Bigr\rangle 
  \underset{N\to\infty}{\longrightarrow}
  \Bigl\langle
    \bigl|\Tr P_{C_1}(U)\bigr|^2
  \Bigr\rangle
  \Bigl\langle
    \bigl|\Tr P_{C_2}(U)\bigr|^2
  \Bigr\rangle\,.
  \label{eq:factorization}
\end{align}
The proof is essentially the same as the proof of \eqref{eq:PCl}, 
or the proof of the large $N$ decomposition of the expectation values of Wilson loops in QCD \cite{migdal1980properties}
(see also \cite{makeenko2002methods} and references therein).

We assume that $P_{C_1}(U)$ and $P_{C_2}(U)$ share a unitary matrix $U$, otherwise \eqref{eq:factorization} is trivial.
We express $\Tr P_{C_1}(U)$ and $\Tr P_{C_2}(U)$ as
\begin{align}
  \Tr P_{C_1}(U) = \Tr \left(U A_1 U A_2 \cdots U A_r\right), \quad
  \Tr P_{C_2}(U) = \Tr \left(U B_1 U B_2 \cdots U B_s\right)\,,
  \nn
\end{align}
as same as \eqref{eq:UinP},
where we have ignored the irrelevant signatures for simplicity.
Then the left-hand side of \eqref{eq:factorization} is written as
\begin{align}
  \Bigl\langle
    &\bigl|\Tr P_{C_1}(U)
    \Tr P_{C_2}(U)\bigr|^2
  \Bigr\rangle \nn \\
  & = \prod_{\underset{U\ne U_e}{e\in E}} dU_e
   \frac{1}{N^{sr}}\Bigl\{  \sum_{\sigma\in S_{r+s}}
   \Bigl(
  \delta_{i_1,i'_{\sigma(1)}}\cdots \delta_{i_n,i'_{\sigma(r+s)}}
  \delta_{j_1,j'_{\sigma(1)}}\cdots \delta_{j_n,j'_{\sigma(r+s)}}\nn \\
  &\hspace{3cm}\times
  (A_1)_{j_1i_2}\cdots(A_r)_{j_r i_1}
  (B_1)_{j_{r+1}i_{r+2}}\cdots(B_s)_{j_{r+s} i_{r+1}} \nn \\
  &\hspace{3cm}\times
  (A^\dagger_r)_{i'_1j'_r}\cdots(A^\dagger_1)_{i'_2j'_1}
  (B^\dagger_r)_{i'_{r+1}j'_{r+s}}\cdots(B^\dagger_1)_{i'_{r+2} j'_{r+1}}\Bigr)
  + {\cal O}(1/N)\Bigr\}\,.
  \nn
\end{align}
From the same discussion as evaluating \eqref{eq:expandPC},
the leading contribution is $N^{r+s}$ which comes from
$\prod_{i=1}^r \Tr(A_iA_i^\dagger)\times \prod_{j=1}^s\Tr(B_jB_j^\dagger)$
and the other contributions disappear at large $N$.
In this case, all of such terms are included in
the summation of $\sigma\in S_r\times S_s \subset S_{r+s}$ which gives
$\Bigl\langle
  \left|\Tr P_{C_1}(U)\right|^2
\Bigr\rangle
\Bigl\langle
  \left|\Tr P_{C_2}(U)\right|^2
\Bigr\rangle$.
This means that all the terms that mix
$\Tr P_{C_1}(U)$ and $\Tr P_{C_2}(U)$ in the integral
vanish at large $N$.
Therefore we can conclude the decomposition \eqref{eq:factorization} in the large $N$ limit.

\subsubsection*{\underline{\it Integral of general Wilson loops}}

Combining \eqref{eq:PCl} and \eqref{eq:factorization},
we see
\[
  \Bigl\langle
    \bigl|\Tr P_{C}(U)^l\bigr|^{2m}\,
  \Bigr\rangle
  \underset{N\to\infty} \longrightarrow 
   l^m m!\,, 
\]
where the factor $m!$ comes from the number of ways to
choose $m$ pairs of $\Tr P_C(U)^l$ and $\Tr P_C(U)^{\dagger l}$
in $\left|\Tr P_{C}(U)^l\right|^{2m}$.

In general, we can associate a partition 
$\lambda_C=(l_1^{m_1} l_2^{m_2}\cdots)$
to a chiral primitive cycle $C$ and consider the quantity, 
\[
  \Upsilon_{\lambda_C}(P_C(U)) \equiv \prod_{i}
  (\Tr P_C(U)^{l_i})^{m_i}\,.
\]
We can then evaluate the integral of a product of
general Wilson loops at large $N$ as
\begin{align}
\Bigl\langle \prod_{C\in \Pi^+} \Upsilon_{\lambda_C}(P_C(U)) \Bigr\rangle
  \underset{N\to\infty} \longrightarrow 
  \prod_{C\in \Pi^+} z_{\lambda_{C}}\,, 
  \label{eq:asifDS}
\end{align}
where $\Pi^+$ is the set of the equivalence classes of 
the chiral primitive cycles in $G$,
and $z_{\lambda_C}$ is given by \eqref{eq:zlambda}.
Comparing this result to the formula
\eqref{eq:DS formula}, we see that we can treat
the Wilson loops corresponding to the chiral primitive cycles
as if independent variables of the unitary integral
on the graph in the large $N$ limit.

\subsection{Evaluation of $Z_G(q)$ at large $N$}

Let us then evaluate the matrix integral $Z_G(q)$ given in \eqref{eq:Zq}
at large $N$ based on the properties we discussed so far.

Using the fact that $Z_G(q)$ counts all the Wilson loops with their length as weights and a reduced cycle is expressed as a power of a primitive cycle,
we can rewrite $Z_G(q)$ as
\begin{align}
  Z_G(q) &= \int \prod_{e\in E} dU_e\,
  e^{ \frac{1}{2}\sum_{n=1}^\infty \frac{q^n}{n}
  \sum_{C\in R_n}{|\Tr P_{C}(U)|^2}
    } \nn \\
    &=
    \Bigl\langle
    \exp\Bigl(
        \frac{1}{2}\sum_{n=1}^\infty \frac{q^n}{n}
        \sum_{d | n} \sum_{C\in \Pi_{d}^+} 2d
        \bigl|\Tr P_C(U)^{\frac{n}{d}}\bigr|^2
      \Bigr)
    \Bigr\rangle \nn \\
    &=
    \Bigl\langle
    \exp\Bigl(
      \sum_{n=1}^\infty \sum_{d | n} \sum_{C\in \Pi_d^+}
      \frac{q^n}{n/d}
      \bigl|\Tr P_C(U)^{\frac{n}{d}}\bigr|^2
    \Bigr)
    \Bigr\rangle \nn \\
    &=
    \Bigl\langle
    \prod_{k=1}^\infty \prod_{C\in \Pi_{k}^+}
    \exp\Bigl(
      \sum_{m=1}^\infty
      \frac{q^{km}}{m}
      \bigl|\Tr P_C(U)^{m}\bigr|^2
    \Bigr)
    \Bigr\rangle\nn \\
    & \underset{N\to\infty} \longrightarrow 
    \prod_{k=1}^\infty \prod_{C\in \Pi_{k}^+}
    \Bigl\langle
    \exp\Bigl(
      \sum_{m=1}^\infty
      \frac{q^{km}}{m}
      \bigl|\Tr P_C(U)^{m}\bigr|^2
    \Bigr)
    \Bigr\rangle\,,
\nn
\end{align}
where we have used the large $N$ decomposition in the last line.

Using the general relation,
\begin{align}
  \exp\Bigl(\sum_{m=1}^\infty \frac{x^m}{m} a_m \Bigr)
  = \sum_{n=0}^\infty x^n
    \sum_{\lambda=(l_1^{m_1}\cdots l_r^{m_r})\vdash n}
    \frac{a_{l_1}^{m_1}\cdots a_{l_r}^{m_r}}{z_\lambda}\,,
\nn
 \end{align}
 we can evaluate
 \begin{align}
   \Bigl\langle
   \exp\Bigl(
     &\sum_{m=1}^\infty
     \frac{q^{km}}{m}
     \bigl|\Tr P_C(U)^{m}\bigr|^2
   \Bigr)
   \Bigr\rangle \nn \\
   &=
   \sum_{n=0}^\infty q^{kn}
   \sum_{\lambda\vdash n}
   \frac{1}{z_\lambda}
   \Bigl\langle
     \bigl|(\Tr P_{C}(U)^{l_1})^{m_1} \cdots (\Tr P_{C}(U)^{l_{r}})^{m_{r}}\bigr|^2
   \Bigr\rangle 
   = \sum_{n=0}^\infty q^{kn}p(n) 
   = \prod_{i=1}^\infty \frac{1}{1-q^{ik}}\,,
 \nn
 \end{align}
where we have used \eqref{eq:asifDS},
$p(n)$ is the number of the partition of $n$,
and we have used the fact that $\prod_{i=1}^\infty(1-q^i)$
is the generating function of $p(n)$. 
Then we conclude that $Z_G(q)$ is written as an infinity product of (the square root of) the Ihara zeta function:
\begin{align}
  Z_G(q) 
  \underset{N\to\infty} \longrightarrow 
   \prod_{k=1}^\infty \prod_{i=1}^\infty
  \frac{1}{(1-q^{ik})^{\ell_k}}
  = \prod_{i=1}^\infty \zeta_G(q^i)^{\frac{1}{2}}\,.
  \label{eq:final result}
\end{align}
This means that the tuned gKM model can be solved exactly in the large $N$ limit. 
We see that the large $N$ limit of \eqref{eq:ZC1 result0} for $|q|<1$ 
becomes \eqref{eq:final result} as expected. 


The key to the derivation of the partition function \eqref{eq:final result} in the large $N$ limit is that the integral of the unitary matrix in \eqref{eq:Zq} is completely separated into the primitive cycles. The same is true for the expression \eqref{eq:dual2} obtained by performing the $U$ integral first. 
In fact, in \cite{migdal1993exact}, an equation satisfied by the eigenvalue distribution of the scalar fields of the original KM model has been derived through evaluating the HCIZ integral in the large $N$ limit. 
Furthermore, an exact solution of this equation was obtained in \cite{gross1992some} where the eigenvalue distribution of the scalar field becomes semi-circle, regardless of the dimension of the square lattice.
Although our model cannot be compared directly with this exact solution of the original KM model because the parameters are adjusted so that the Ihara zeta function is realized, 
it is interesting that the scalar fields on the any dimensional square lattice behave in the same way as those on the one-dimensional square lattice.
The one dimensional square lattice is nothing but the cycle graph discussed in Sec.~\ref{sec:ZC1}, and the partition function in large $N$ limit becomes $Z_{C_n}(q) \underset{N\to\infty} \longrightarrow  \prod_{i=1}^\infty (1-q^{ni})^{-1}$ from \eqref{eq:ZCn} and \eqref{eq:ZC1 result0}.
Combining this observation to \eqref{eq:final result}, 
we see that the partition function of the gKM model on an arbitrary graph $G$ can be rewritten as a product of the partition function of the model on the cycle graphs in the large $N$ limit: 
\begin{align}
  Z_G(q) 
  \underset{N\to\infty} \longrightarrow \prod_{i=1}^\infty \prod_{[C]\in \Pi^+} \frac{1}{1-q^{\ell(C)i}}
  =\prod_{[C]\in \Pi^+} Z_{C_{\ell(C)}}(q)\,. \nn
\end{align}
Therefore, the gKM model in the large $N$ limit on any graph $G$ can be expected to behave similarly to the model on the cycle graph in general, which supports the results obtained with \cite{gross1992some} in an extended way.


As another comment, it is suggestive that the expression \eqref{eq:final result} can be written as
\[
\begin{split}
\prod_{i=1}^\infty\zeta_G(q^i)^{\frac{1}{2}}
&= \exp\left( \frac{1}{2}\sum_{i=1}^\infty \sum_{k=1}^\infty \frac{N_k}{k} q^{ik} 
\right)\\
&=\exp\left( \frac{1}{2}\sum_{k=1}^\infty
\frac{
\sum_{d|k} d N_{k/d} }{k}
q^k 
\right)
\,,
\end{split}
\]
using $N_k$ in (\ref{eq:Ihara zeta as generating function}).
Regarding it as an Ihara zeta function, the number of the reduced cycles of length $k$ in the corresponding graph is
counted
by $N_k' =\sum_{d|k} d N_{k/d}$
and the free energy has
the exact coupling expansion in $q$,
in the large $N$ limit of the gKM.
The meaning of this expression
will be discussed separately.


\subsection{Finite $N$ effect}
We make a comment on $Z_G(q)$ on finite $N$. 
Looking at the results \eqref{eq:ZC1 result} and \eqref{eq:final result}, 
one may think that \eqref{eq:final result} might hold 
even for finite $N$, but it is not the case in general. 

As an example, we consider the graph showed in Fig.\ref{fig:burtterfly}, 
which is the simplified version of the double triangle graph shown 
in Appendix \ref{Ihara examples}. 
\begin{figure}[H]
  \begin{center}
  \includegraphics[scale=0.6]{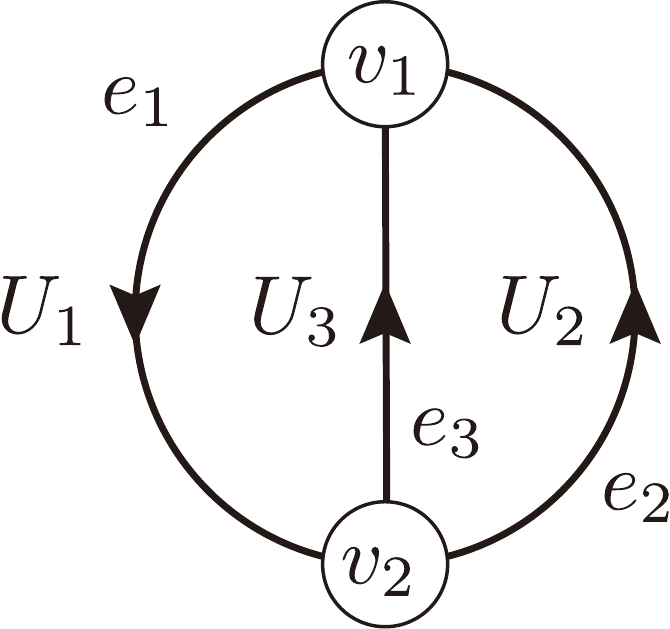}
\end{center}
\caption{An example of the graph.}
\label{fig:burtterfly}
\end{figure}
Writing $W_k(U) \equiv \Tr W_U^k$, we can easily show $W_{2i-1}(U)=0$ ($i\in\N$)
for this graph and the expansion of the extended zeta function is expanded as  
\begin{align}
  \zeta_G(q;U)^\frac{1}{2}= 1 
  &+ q^2 \left[\frac{W_2(U)}{2}\right]
  + q^4 \left[\frac{W_4(U)}{8}+\frac{W_2(U)^2}{32}\right] \nn \\
  &+ q^6 \left[\frac{W_6(U)}{12}+\frac{W_2(U)W_4(U)}{32}+\frac{W_2(U)^3}{384}\right]
  + {\cal O}(q^8)\,,
 \nn
\end{align}
with
\begin{align}
  W_2(U) = 4 \Bigl( &|\Tr(U_1U_2)|^2 + |\Tr(U_1U_3)|^2 + |\Tr(U_2U_3^\dagger)|^2 \Bigr)\,, \nn \\
  W_4(U) = 4 \Bigl( 
    &|\Tr(U_1U_2U_1U_2)|^2 +  |\Tr(U_1U_3U_1U_3)|^2 + |\Tr(U_2U_3^\dagger U_2U_3^\dagger)|^2 \nn \\
    &+2|\Tr(U_1U_2U_1U_3)|^2  +2|\Tr(U_1U_2U_3^\dagger U_2)|^2  +2|\Tr(U_1U_3U_2^\dagger U_3)|^2\,, \nn \\
  W_6(U) = 4\Bigl(
  &|\Tr(U_1U_2U_1U_2U_1U_2)|^2 + |\Tr(U_1U_3U_1U_3U_1U_3)|^2 + |\Tr(U_2U_3^\dagger U_2U_3^\dagger U_2U_3^\dagger)|^2 \nn \\
  &
  +3|\Tr(U_1U_2U_1U_2U_1U_3)|^2
  +3|\Tr(U_2U_1U_2U_1U_2U_3^\dagger)|^2
  +3|\Tr(U_1U_3U_1U_3U_1U_2)|^2 \nn \\
  &+3|\Tr(U_1U_2U_3^\dagger U_2U_3^\dagger U_2)|^2
  +3|\Tr(U_2U_1U_3U_1U_2U_3^\dagger)|^2
  +3|\Tr(U_3U_1U_2U_3^\dagger U_1^\dagger U_2^\dagger)|^2 \nn \\
  &+3|\Tr(U_2U_1U_2U_3^\dagger U_1^\dagger U_3^\dagger )|^2
  +3|\Tr(U_3U_1U_3U_2^\dagger U_3U_2^\dagger )|^2
  +3|\Tr(U_3U_1U_2U_1U_3U_2^\dagger)|^2 \nn \\
  &+3|\Tr(U_3U_1U_3U_1U_3U_2^\dagger)|^2
  \Bigr)\,.
\nn
\end{align}
After a straightforward calculation using \eqref{eq:basic int} and \eqref{eq:DS formula}, we obtain 
\begin{align}
\begin{split}
&\int dU_1 dU_2 dU_3 \biggl(\frac{W_2(U)}{2} \biggr)
= 3\,,  \\
&\int dU_1 dU_2 dU_3 \biggl( \frac{W_4(U)}{8}+\frac{W_2(U)^2}{32} \biggr)
= 12\,, \\ 
&\int dU_1 dU_2 dU_3 \biggl( \frac{W_6(U)}{12}+\frac{W_2(U)W_4(U)}{32}+\frac{W_2(U)^3}{384} \biggr)
= 41 + \frac{8}{N^2-1}\,,
\end{split}
\nn
\end{align}
for $N\ge 3$.
Comparing it to the expansion for $N\ge 3$, 
\[
  \prod_{i=1}^N \zeta_G(q^i)^\frac{1}{2} = 1 + 3q^2 +12q^4 +41q^6 + O(q^8)\,, 
  \quad (N\ge 3)
\]
it is obvious that $\int \prod_e dU_e \, \zeta_G(q;U)^\frac{1}{2} \ne \prod_{i=1}^N \zeta_G(q^i)^\frac{1}{2}$ at finite $N$. 
This is of course expected from the analysis made in Sect.~\ref{sec:decomp}.

\section{Conclusion and Discussion}
\label{sec:Conclusion and Discussion}
In this paper, as an extension of the Ihara zeta function associated with graphs,
we proposed a function that generates all non-collapsing Wilson loops with their lengths as weights.
We also proposed the Kazakov-Migdal model on graphs (gKM model) and showed that, when the parameters of the model are appropriately tuned, the partition function is represented by the unitary matrix integral of the extended Ihara zeta function. 
Since the KM model has both Hermitian matrices and unitary matrices, there are two different ways to represent the same partition function, depending on which integral is performed first.
Applying this duality to the tuned gKM model of a cyclic graph (single polygon), 
we exactly evaluated the unitary matrix integral 
$\int dU \exp\lrpar{\sum_{n=1}^\infty \frac{q^n}{n} |\Tr U^n|^2}$
at finite $N$.
We also discussed a possibility to analytically connect this integral to the region $|q|>1$.
We further showed that the partition function of the tuned gKM model 
corresponding to any graph can be computed exactly at large $N$ 
and expressed as the infinite product of the square roots of the Ihara zeta functions. 

The tuned gKM model is the minimal integral to count all the Wilson loops by unitary matrix integral 
and this is exactly the reason why the partition function of this model is expressed as a product of the {\em square roots} of the Ihara zeta function rather than the zeta function itself. 
The point is that an integral of a unitary matrix $U$ vanishes unless the integrand contains equal numbers of $U_{ij}$ and $U_{ij}^*$. 
For this reason, the simple integration of a Wilson loop 
$\int \prod_{e\in E} dU_e \Tr P_C(U)$ on the graph generally vanishes, except for such special Wilson loops that happen to contain equal numbers of $U_{ij}$ and $U_{ij}^*$. 
Therefore, in order that the $U$-integral of a Wilson loop always produces a non-zero value, we need to pair a Wilson loop with its Hermitian conjugate and evaluate the quantity $\int \prod_{e\in E} dU_e |\Tr P_C(U)|^2$.
From the standpoint of the cycles on the graph, 
the Wilson loop and its Hermitian conjugate 
correspond to a reduced cycle $C$ and its inverse cycle $C^{-1}$, respectively. 
Then, if we want to count up all the Wilson loops by unitary matrix integral, 
we have only to count up exactly half of the reduced cycles, 
which is nothing but the square root of the Ihara zeta function.
This suggests that the gKM model on an arbitrary graph becomes effectively a direct product of the models on cycle graphs (one-dimensional lattices) in the large $N$ limit. 
This is expected to be consistent with the result obtained in \cite{gross1992some} that the exact solution of the original KM model exhibits universal behavior regardless of the lattice dimension.

As discussed in Sect.~\ref{sec:ZC1}, 
in evaluating the integral 
$\int dU \exp\lrpar{\sum_{n=1}^\infty \frac{q^n}{n} |\Tr U^n|^2}$, 
we did not integrate the unitary matrix directly, 
but rather performed the Hermitian matrix integral \eqref{eq:I1_med}
using the duality of the tuned gKM model.
Interestingly enough, 
this Hermitian matrix integral is defined not only for $|q|<1$ but also for $|q|>1$ as in \eqref{eq:ZC1 result0}. 
On the other hand, the tuned gKM model is defined only for $|q|<1$ 
since otherwise $\beta$ becomes negative. 
It is also clear from the fact that the partition function at large $N$ is represented by the Ihara zeta function, whose infinite product representation is defined only for $q$ with sufficiently small $|q|$.
Therefore, we would normally expect that only the result \eqref{eq:ZC1 result0} in $|q|<1$ would be meaningful. 
However, as we show explicitly in Appendix \ref{app:q1}, 
we can perform the unitary matrix integration directly for small $N$ 
by using Cauchy's integral theorem both in the regions $|q|<1$ and $|q|>1$, 
and the results are identical with \eqref{eq:ZC1 result0} 
and \eqref{eq:ZC1 result qlt1}, respectively.
This suggests that this unitary matrix integral can be analytically connected to the region of $|q|>1$. 
This also opens an possibility for the tuned gKM model to be extended to the region $|q|>1$.


It is also interesting to think of the tuned gKM model as 
an induced QCD as discussed in the original KM model. 
The original KM model does indeed add up all Wilson loops, 
but it includes also the contributions of backtrackings 
and the length of those Wilson loops does not match the power of $q$.
By applying the parameter setting introduced in this paper, 
the relation between the length of the Wilson loop and the power of $q$ is established, which is expected to make the connection to the gauge theory easier. 
Furthermore, 
although we mainly concentrate on $Z_G(q)$ in the partition function \eqref{eq:Zq}
in this paper, 
the partition function also includes the factor $\alpha^{-\frac{n_V N^2}{2}}$. 
In order to evaluate the large $N$ behavior of the model, 
we would have to control the partition function by carefully 
tuning the value of the factor $\alpha$. 
We may then see the large $N$ behavior of QCD explicitly through 
the exact expression of the partition function obtained in this paper.

In \cite{Matsuura:2014nga}, it is shown that HCIZ integral and KM model
can be embedded into a supersymmetric model on the graph,
where the matrix integral is evaluated as a result of the Duistermaat-Heckman localization formula (fixed point theorem).
We expect that our (tuned) gKM model can also be embedded into a supersymmetric model 
and the related Ihara zeta function can be understood as
a vev of the supersymmetric Wilson loops.
On the other hand, the partition function of the supersymmetric gauge theory,
including the quiver quantum mechanics (see e.g.~\cite{Ohta:2014ria,Ohta:2015fpe}),
gives various kinds of supersymmetric indices, 
which count the number of
the gauge invariant operators (chiral rings) or BPS bound states.
So it is interesting to find a relation between the BPS state counting
and Ihara zeta function via the supersymmetric quiver gauge theories
on the graph. Since the supersymmetric quiver quantum mechanics
also counts the BPS bound states of the multi-centered black holes
in the dual gravitational (closed string) picture \cite{Denef:2002ru},
the Ihara zeta function, involving graph theory as well,
has the potential to shed new light on gravity and string theory.

\section*{Acknowledgments}
We would especially like to thank S.~Nishigaki for suggesting the relation between gauge theory on graphs and the Ihara zeta function.
This work is supported in part
by Grant-in-Aid for Scientific Research (KAKENHI) (C) Grant Number
17K05422 (K.~O.)
and  Grant-in-Aid for Scientific Research (KAKENHI) (C), Grant Number 20K03934 (S.~M.).

\appendix

\section{Some examples of the Ihara zeta function}
\label{Ihara examples}

\subsection{Cycle graph}

The cycle graph $C_n$ is shown in Fig.~\ref{Figure of the cycle graph} (a).
$C_n$ has $n$ vertices and edges, namely $n_V=n_E=n$.
The adjacency matrix is an $n\times n$ matrix and given by
\[
A_{C_n} = \begin{pmatrix}
0 & 1 & \cdots & 0 & 1\\
1 & 0 & \cdots & 0 & 0\\
\vdots & \vdots & \ddots & \vdots & \vdots\\
0 & 0 & \cdots & 0 & 1\\
1 & 0 & \cdots & 1 & 0
\end{pmatrix}\,,
\]
and the degree matrix is given by $D_{C_n}=\diag(2,2,\ldots,2)$.
Then the Ihara zeta function is given by an inverse of the 
determinant of the following matrix
\[
{\bf 1}_n -q A_{C_n}+q^2(D_{C_n}-{\bf 1}_n)=
 \begin{pmatrix}
1+q^2 & -q & \cdots & 0 & -q\\
-q & 1+q^2 & \cdots & 0 & 0\\
\vdots & \vdots & \ddots & \vdots & \vdots\\
0 & 0 & \cdots & 1+q^2 & -q\\
-q & 0 & \cdots & -q & 1+q^2
\end{pmatrix}\,,
\]
which is nothing but the $q$-deformed Cartan matrix of the affine Lie algebra $\hat{A}_{n-1}$ (the $C_n$ graph
is the Dynkin diagram of $\hat{A}_{n-1}$). Thus we get
\be
\begin{split}
\zeta_{C_n}(q)
&= \frac{1}{\det\left({\bf 1}_n -q A_{C_n}+q^2(D_{C_n}-{\bf 1}_n)\right)}\\
&=\frac{1}{(1-q^n)^2}\,.
\nn
\end{split}
\ee
The square in this expression means that there are a primitive cycle of the length $n$ and its inverse
(anti-clockwise and clockwise cycle). So the square root of the Ihara zeta function
picks up the primitive cycle of one direction only, 
which is the chiral primitive cycle.

The logarithm of the Ihara zeta function has the expansion, 
\[
\log \zeta_{C_n}(q) = 2\sum_{m=1}^\infty \frac{q^{nm}}{m}
=\sum_{m=1}^\infty \frac{2n}{nm}q^{nm}\,.
\]
This means that $N_k = 2n$ iff $k\equiv 0 \pmod{n}$
and then all the reduced cycles are expressed by the power of the
one-directed primitive cycle of the length $n$.

\subsection{Double triangle}

\begin{figure}[H]
\begin{center}
\includegraphics[scale=0.5]{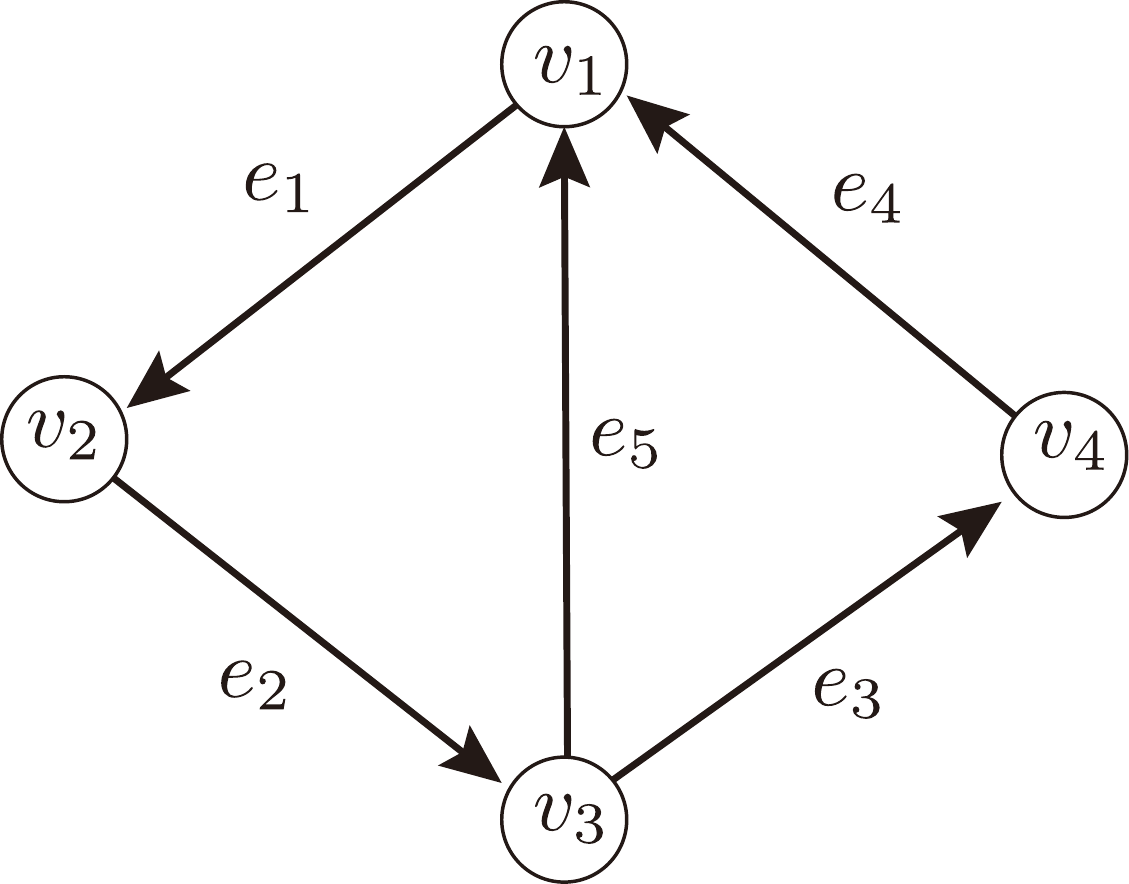}
\end{center}
\caption{This graph has four vertices and five edges. Two triangles are glued by the edge $e_5$.}
\label{double triangle graph}
\end{figure}

Here we consider the graph depicted in Fig.~\ref{double triangle graph}.
This graph contains two triangles, so we call this graph the double triangle (DT),
which is also obtained by removing one edge from the complete graph $K_4$.
DT has four vertices and five edges ($n_V=4$ and $n_E=5$).
There are infinitely many primitive cycles on DT,
since there are infinite ways to go around the combination of the different cycles.
So the original definition of the Ihara
zeta function becomes an infinite product over the primitive cycles.

However, Ihara's theorem says that the Ihara zeta function is expressed by a determinant made from
the $4\times 4$ adjacency matrix of DT
\[
A_{\rm DT}=\begin{pmatrix}
0 & 1 & 1 & 1\\
1 & 0 & 1 & 0\\
1 & 1 & 0 & 1\\
1 & 0 & 1 & 0
\end{pmatrix}\,,
\]
and the degree matrix $D_{\rm DT}=\diag(3,2,3,2)$.
We obtain a rational function of $q$
\[
\begin{split}
\zeta_{\rm DT}(q) &= \frac{1}{(1-q^2)\det\left({\bf 1}_4 - qA_{\rm DT} + q^2 (D_{\rm DT}-{\bf 1}_4)\right)} \\
&=\frac{1}{(1-q^4)(1+q^2-2q^3)(1-q^2-2q^3)}\,.
\end{split}
\]
The logarithm of the Ihara zeta function has the following series expansion
\[
\log \zeta_{\rm DT}(q)
=4 q^3+2 q^4+4 q^6+4 q^7+q^8+\frac{16 q^9}{3}+12 q^{10}+4 q^{11}+\frac{26 q^{12}}{3}+\cdots.
\]
So we see
\[
\begin{split}
&N_3 = 12,\ 
N_4 = 8,\ 
N_6 = 24,\ 
N_7 = 28,\ 
N_8 = 8,\\
&N_9 = 48,\ 
N_{10} = 120,\ 
N_{11} = 44,\ 
N_{12} = 104,\ \cdots,
\end{split}
\]
and the others are zero.
Note that $N_k$ are even numbers,
since the equivalent class of the reduced cycle always contains the cycle
of both directions.
For example, $N_3/2=3+3$ stands for the number of two chiral primitive cycles
with length three on each triangle,
$N_4/2=4$ stands for the number of chiral primitive cycles
around the outer rhombus with length four,
$N_6/2=3+3+6$ stands for two squares of length three chiral primitive cycles on each triangle,
and a length six product of two chiral primitive cycles (triangles),
and so on.

\subsection{Tetrahedron}

\begin{figure}[H]
\begin{center}
\subcaptionbox{Tetrahedron graph}[.45\textwidth]{
\includegraphics[scale=0.5]{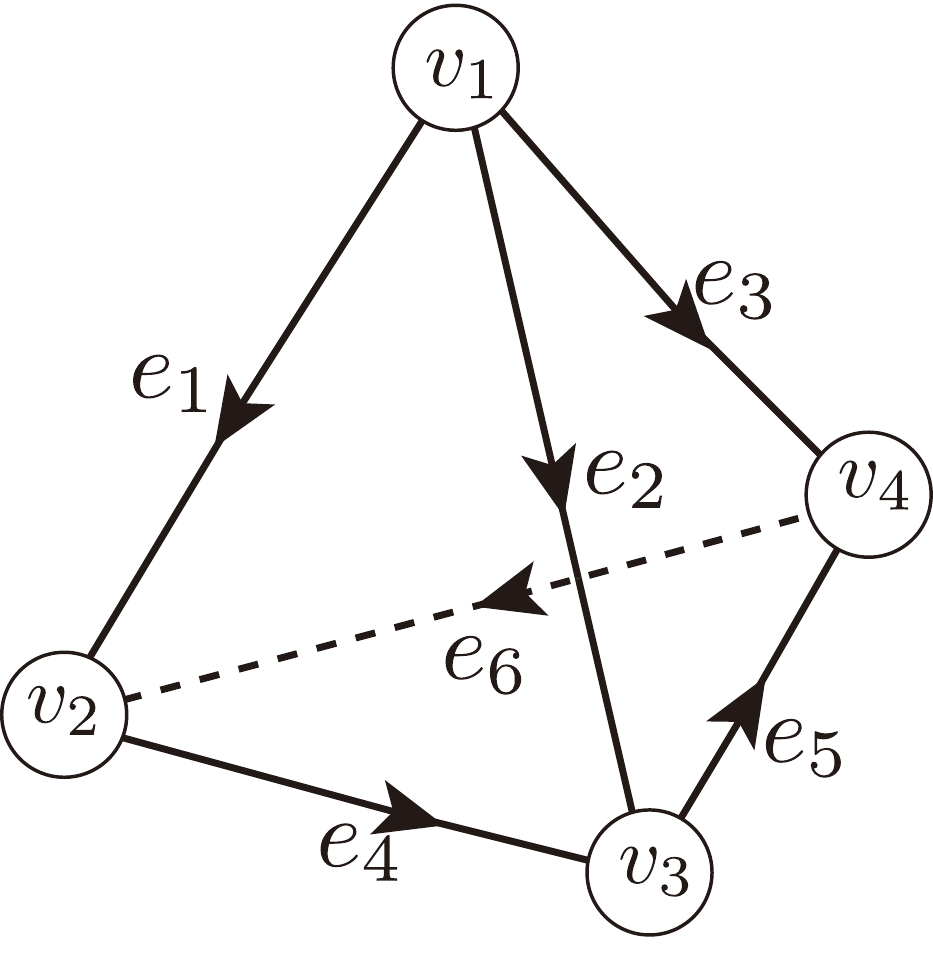}
}
\subcaptionbox{Equivalent complete graph $K_4$}[.45\textwidth]{
\includegraphics[scale=0.5]{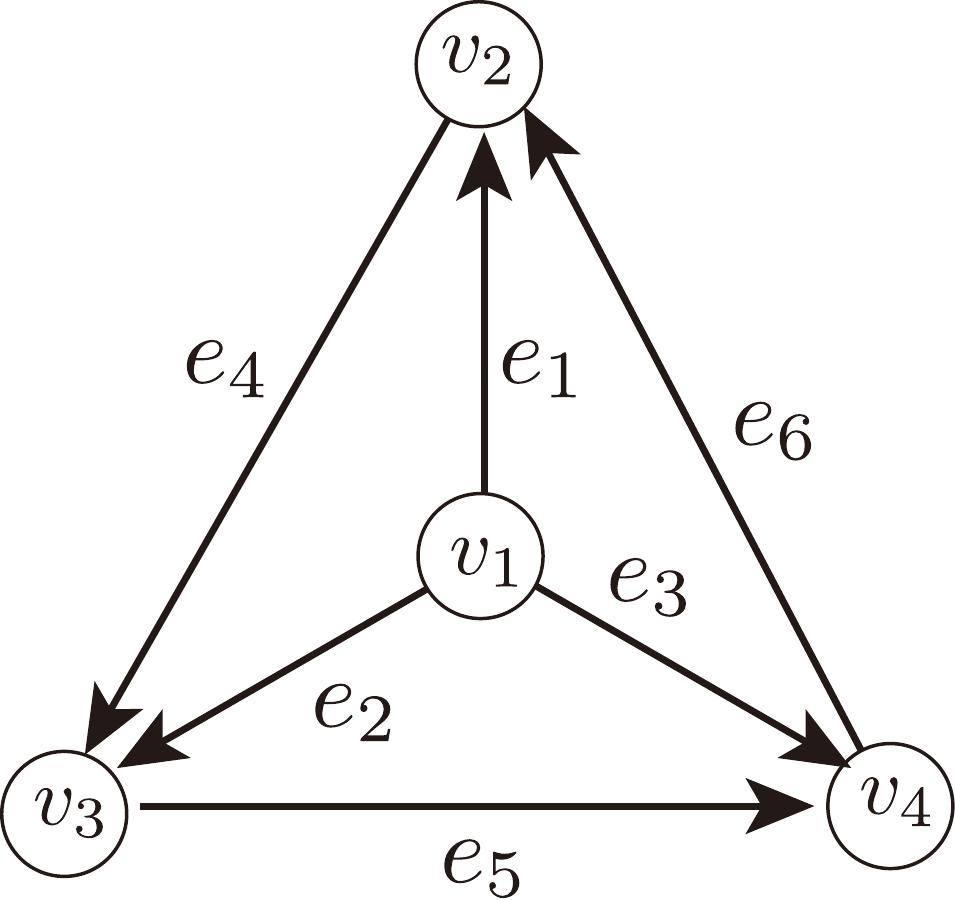}
}
\end{center}
\caption{This graph has four vertices and six edges and forms a tetrahedron, where
four triangles are gluing with each other.
The three-dimensional tetrahedron graph is equivalent to a planner complete graph $K_4$,
since the graph has only the information on the connections between vertices.}
\label{tetrahedron graph}
\end{figure}

If we would like to consider the Wilson loops
of the gauge theory defined on the surface of the polyhedra
\cite{Ohta:2021tmk},
it is useful to consider the polyhedra as graphs. 
For example, let us consider the tetrahedron.
If we give connections between vertices by edges of the tetrahedron,
we obtain a graph in the three-dimensional picture depicted in Fig.~\ref{tetrahedron graph} (a).
The tetrahedron graph is equivalent to the planner graph shown in
Fig.~\ref{tetrahedron graph} (b) and (c).
This graph is also called the complete graph $K_4$, 
which has four vertices and six edges ($n_V=4$ and $n_E=6$).

The adjacency matrix of the complete graph $K_4$ is given by
\[
A_{K_4} = \begin{pmatrix}
0 & 1 & 1 & 1\\
1 & 0 & 1 & 1\\
1 & 1 & 0 & 1\\
1 & 1 & 1 & 0
\end{pmatrix},
\]
and the degree matrix is $D_{K_4}=\diag(3,3,3,3)$.
Then we can obtain the Ihara zeta function,
\be
\begin{split}
\zeta_{K_4}(q) &= \frac{1}{(1-q^2)^2\det\left({\bf 1}_4 - qA_{K_4} + q^2 (D_{K_4}-{\bf 1}_4)\right)} \\
&=\frac{1}{(1-q) (1-2 q)(1-q^2)^2 (1+q+2 q^2)^3}\,.
\end{split}
\nn
\ee
The logarithm of the Ihara zeta function for $K_4$ has the series expansion, 
\[
\log \zeta_{K_4}(q)
=8 q^3+6 q^4+16 q^6+24 q^7+21 q^8+\frac{176 q^9}{3}+120 q^{10}+168 q^{11}+330 q^{12}+\cdots.
\]
From this, we have
\[
\begin{split}
&N_3 = 24,\ 
N_4 = 24,\ 
N_6 = 96,\ 
N_7 = 168,\ 
N_8 = 168,\\
&N_9 = 528,\ 
N_{10} = 1200,\ 
N_{11} = 1848,\ 
N_{12} = 3960,\ \cdots.
\end{split}
\]
The number of the reduced cycles increases more rapidly than the case of the DT. 

\section{Formulas for symmetric polynomials}
\label{app:proof}
In this appendix, we give a proof of the formulas \eqref{eq:th1-2} and \eqref{eq:th1}%
\footnote{See e.g.~\cite{macdonald1998symmetric} for more details of the symmetric polynomials.}.

We first define the homogeneous symmetric function associated to $n\in\N$ by
\[
  h_n(\bfx) \equiv \sum_{\substack{i_j\in\{0,1,2,\cdots\} \\\sum_{j} i_j = n}} \prod_{j=1}^\infty x_{i_j}\,,
  \label{eq:homSF}
\]
where $\bfx=\{x_i|i\in\N\}$.
It is easy to see that the generating function of the homogeneous symmetric functions is given by
\[
  H(t) = \prod_{i=1}^\infty \frac{1}{1-tx_i} = \sum_{n=0}^\infty t^n h_n(\bfx) \,.
\]

Here we transform $H(t)$ as
\begin{align}
  H(t)&=\sum_{n=0}^\infty t^n h_n(\bfx) = \prod_{i=1}^\infty \frac{1}{1-tx_i}
  = \exp\left( -\sum_{i=1}^\infty \log(1-tx_i)\right)
  = \exp\left( \sum_{i=1}^\infty \sum_{k=1}^\infty \frac{t^kx_i^k}{k} \right) \nn \\
  &= \exp\left( \sum_{l=1}^\infty \frac{t^l p_l(\bfx)}{l} \right)
  = \prod_{l=1}^\infty \exp\left( \frac{t^l p_l(\bfx)}{l} \right)
  = \prod_{l=1}^\infty \sum_{m_l=0}^\infty \frac{p_l(\bfx)^{m_l}}{m_{l}! l^{m_l}} t^{lm_l}
  = \sum_{n=0}^\infty t^n \sum_{\lambda \vdash n} \frac{p_\lambda(\bfx)}{z_\lambda}\,,
\nn
\end{align}
where $p_n(\bfx)$ is called the power sum symmetric function, 
\[
  p_n(\bfx) \equiv \sum_{i}^\infty x_i^n\,, 
\]
and
\[
  p_\lambda(\bfx) \equiv \prod_{l=1}^r p_{\lambda_l}(\bfx)\,,
\]
for a partition $\lambda=(\lambda_1,\cdots,\lambda_r)$ of $N$ which satisfies
\[
\lambda_1\ge\cdots \ge \lambda_r,  \quad  |\lambda|\equiv \sum_{i=1}^r \lambda_r = N\,.
\]
Comparing the coefficient of $t^n$, we obtain the relation, 
\begin{equation}
  h_n(\bfx) = \sum_{\lambda \vdash N } \frac{p_\lambda (\bfx)}{z_\lambda}\,.
  \label{eq:h by p}
\end{equation}

\begin{figure}[H]
  \begin{center}
  \includegraphics[scale=0.6]{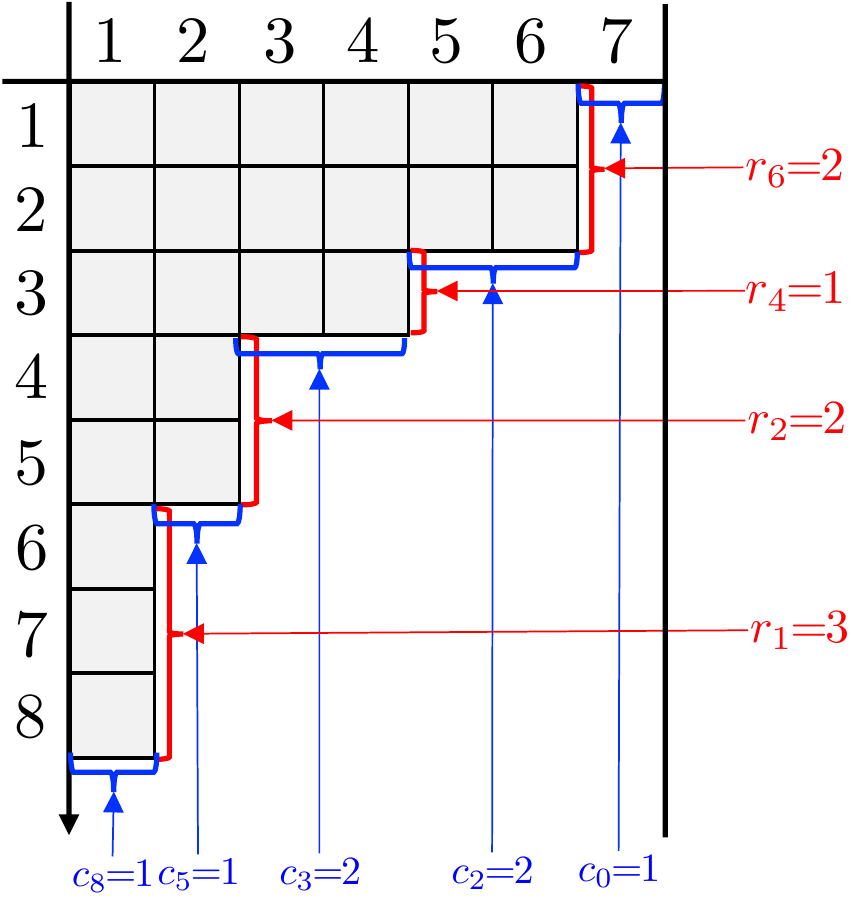}
\end{center}
\caption{An example of Young diagrams whose maximal length of row is $7$.}
\label{fig:YD}
\end{figure}

We next consider Young diagrams whose maximal length of row is $N$. 
The generating function of the number of such Young diagrams is given by
\begin{align}
  P_N(q) &= \prod_{i=1}^N \frac{1}{1-q^i}
  = \sum_{\substack{c_j\in\{0,1,2,\cdots\} \\ \sum_{j=0}^\infty c_j = N}} \prod_{j=0}^\infty q^{j{c_j}}
  \label{eq:PN}
\end{align}
In fact, 
the Taylor series of the middle expression of \eqref{eq:PN} is
\begin{align}
  \prod_{i=1}^N \sum_{r_i=0}^\infty q^{i r_i} = \sum_{r_i=0}^\infty q^{\sum_{i=0}^N i r_i}\,,
\nn
\end{align}
whose exponent of $q$, $\sum_{i=1}^N i r_i$, 
can be regarded as the number of boxes of such Young diagram which has $r_i$ rows of length $i$.
On the other hand, let $c_j$ be the number of columns with length $j$ $(j\in\{0,1,2,\cdots\})$.
Since we are now considering a Young diagram with row lengths of at most $N$,
the sum of $c_j$ must be $N$ (see Fig.~\ref{fig:YD}). 
This means that the right-hand side of \eqref{eq:PN} is also the generating function.

Combining \eqref{eq:h by p} and \eqref{eq:PN}, we obtain \eqref{eq:th1-2} and \eqref{eq:th1}.
From the relation 
\[
\frac{1}{1-q^n}=\sum_{i=0}^\infty q^{ni} = p_n(\bfx)|_{x_i=q^{i-1}}\equiv p_n(q), 
\]
we see
\[
  \sum_{\lambda\vdash N} \frac{1}{z_\lambda}\prod_{i=1}^r\frac{1}{1-q^{\lambda_i} }
  = \sum_{\lambda\vdash N} \frac{p_\lambda(q)}{z_\lambda}
  = h_N(q) \equiv h_N(\bfx)|_{x_i=q^{i-1}}\,,
\]
where we have used \eqref{eq:h by p} in the last equality.
From the definition of the homogeneous symmetric function \eqref{eq:homSF},
we obtain
\[
h_N(q) = \sum_{\substack{c_j\in\{0,1,2,\cdots\} \\\sum_{j=0}^\infty c_j = N}} \prod_{j=0}^\infty q^{jc_j}\,,
\]
which is equal to the right-hand side of \eqref{eq:PN}.
We can then conclude
\[
  \sum_{\lambda\vdash N} \frac{1}{z_\lambda}\prod_{i=1}^r\frac{1}{1-q^{\lambda_i} }
  = P_N(q) = \prod_{i=1}^N \frac{1}{1-q^i}\,, 
\]
which is \eqref{eq:th1}. 
\eqref{eq:th1-2} is obtained as a corollary of this;
\begin{align}
  \sum_{\lambda\vdash N} \frac{1}{z_\lambda}\prod_{i=1}^r\frac{1}{q^{\lambda_i}-1 }
  &= \sum_{\lambda\vdash N} \frac{1}{z_\lambda}\frac{1}{q^N}\prod_{i=1}^r\frac{1}{1-q^{-\lambda_i} } 
  = \frac{1}{q^N} \prod_{i=1}^N \frac{1}{1-q^{-i}}  \nn \\
  &= \frac{q^{N(N+1)/2}}{q^{N}} \prod_{i=1}^N \frac{1}{q^i-1} 
  = {(-1)^N}{q^{N(N-1)/2}}\prod_{i=1}^N \frac{1}{1-q^i}\,.
\nn
\end{align}

\section{Direct computation of $Z_{C_1}(q)$ for finite $N$}
\label{app:q1}
In this appendix, we compute 
\begin{equation}
  Z_{C_1}(q)=  \int dU\,
  e^{ \sum_{n=1}^\infty \frac{q^n}{n}
  |\Tr U^n|^2 }
  \label{eq:ZC1}
\end{equation}
directly using Cauchy's integral theorem. 

To this end, we need the explicit expression of the Haar measure of $U(N)$, 
\begin{align}
  \int dU &= \frac{1}{N!} \prod_{i=1}^N \lrpar{\int_0^{2\pi}\frac{d\theta_i}{2\pi}} \prod_{j<k}
  |e^{i\theta_j}-e^{i\theta_k}|^2 \nn \\
  &= \frac{1}{N!} \prod_{i=1}^N\lrpar{\oint_{|z_i|=1} \frac{dz_i}{2\pi z_i}}
  \prod_{j<k} (z_j-z_k)(z_j^{-1}-z_k^{-1})\,,
\nn
\end{align}
where $z_i=e^{i\theta_i}$ ($i=1,\cdots,N$) are the eigenvalues of $U$. 
We also rewrite the exponent in \eqref{eq:ZC1} as 
\begin{align}
  \sum_{n=1}^\infty \frac{q^n}{n} |\Tr U^n|^2
  &= \sum_{n=1}^\infty \Bigl| \frac{q^n}{n} \sum_{j=1}^\infty e^{in\theta_j} \Bigr|^2 \nn \\
  &= \sum_{n=1}^\infty \frac{q^n}{n} 
     \lrpar{N + \sum_{1\le j < k \le N}(z_j^n z_k^{-n} + z_j^{-n}z_k^n)} \nn \\
  &= N\sum_{n=1}^\infty \frac{q^n}{n} 
     + \sum_{j<k} \sum_{n=1}^\infty \frac{1}{n} \Bigl( (q z_j z_k^{-1})^n + (q z_j^{-1} z_k)^n \Bigr) \nn \\
  &= -\log\Bigl( (1-q)^N\prod_{j<k}(1-qz_j z_k^{-1})(1-q z_j^{-1} z_k) \Bigr)\,. 
\nn
\end{align}
Then \eqref{eq:ZC1} can be written as 
\begin{align}
  Z_{C_1}(q) &= \frac{1}{N! (1-q)^N} 
   \prod_{i=1}^N\lrpar{\oint_{|z_i|=1} \frac{dz_i}{2\pi z_i}}
  \prod_{j<k} \frac{(z_j-z_k)(z_j^{-1}-z_k^{-1})}
  {(1-qz_j z_k^{-1})(1-q z_j^{-1} z_k) }\,.
  \label{eq:ZC1res}
\end{align}

We first assume $|q|<1$. 
The strategy to compute $Z_{C_1}(q)$ is as follows:
Recalling $|z_i|=1$ ($i=1,\cdots,N$), 
the simple poles of $z_N$ in \eqref{eq:ZC1res} inside $|z_N|=1$ are at $z_N=0$ and $z_N=q z_i$ 
($i=1,\cdots,N-1)$. 
From Cauchy's integration theorem, the integration over $z_N$ is obtained 
by summing over all the corresponding residues. 
Then the simple poles of $z_{N-1}$ in the remained function inside of $|z_{N-1}|=1$ 
are at $z_{N-1}=0$ and $z_{N-1}=q^k z_{i}$ with $k=1,2$ and $i=1,\cdots,N-2$. 
If we repeat this procedure for $m$ times, 
the simple poles of $z_{N-m}$ in the remaining function inside of $|z_{N-m}|=1$
are at 
$z_{N-m}=0$ and $z_{N-m}=q^k z_{i}$ with $k=1,\cdots,m$ and $i=1,\cdots,N-m-1$. 
Finally, after repeating this procedure $N-1$ times, 
the remaining function of $z_1$ has only one simple pole at $z_1=0$ 
and we obtain $Z_{C_1}(q)$ from the residue. 

We carried it out by hand for $N=1$ and $N=2$ and by computer (Maple) for $N=3,\cdots,6$. 
The result is 
\[
  Z_{C_1}(q) = \prod_{i=1}^N \frac{1}{1-q^i}\,, 
\]
which is the same with \eqref{eq:ZC1 result}. 

Interestingly, we can perform the same computation for $|q|>1$. 
The strategy is the same for the case of $|q|<1$ but now the residues 
appear at $z_{N-m}=0$ and $z_{N-m}=z_{i}/q^k$ with $k=1,\cdots,m$ and $i=1,\cdots,N-m-1$ after repeating the procedure $m$ times.
Then the result for $N=1,\cdots,6$ is 
\begin{equation}
  Z_{C_1}(q) = \prod_{i=1}^N \frac{1}{q^{i-1}-q^{2i-1}}\,, 
  \label{eq:ZC1qgt1}
\end{equation}
which is again the same with \eqref{eq:ZC1 result}. 

In deriving the expression \eqref{eq:ZC1res}, we have used 
$\log(1-x)=\sum_{n=1}^\infty \frac{x^n}{n}$, 
which is justified only in the region $|x|<1$. 
In this sense, \eqref{eq:ZC1qgt1} is a result outside the range of applicability of the calculation and is essentially meaningless. 
Nevertheless, the results of the direct integration of the unitary matrix and the integration by the Hermitian matrix using the duality agree.
We can then expect that \eqref{eq:ZC1qgt1} is the analytic continuation 
of the integral \eqref{eq:ZC1} to $|q|>1$. 

\bibliographystyle{unsrt}
\bibliography{refs}

\end{document}